\documentclass[a4paper,onecolumn,11pt,accepted=2025-02-17]{quantumarticle}
\pdfoutput=1

\usepackage{graphicx,xcolor}
\usepackage{amsmath,amssymb,amsfonts,amsthm, microtype, mathrsfs,bbm}
\usepackage[colorlinks,allcolors=blue]{hyperref}
\usepackage{braket,mathtools,physics,enumerate}
\usepackage[capitalize]{cleveref}  
\usepackage{cases}
\usepackage{thm-restate}
\newtheorem{theorem}{Theorem}
\newtheorem{definition}[theorem]{Definition}
\newtheorem{proposition}[theorem]{Proposition}
\newtheorem{lemma}[theorem]{Lemma}
\newtheorem{corollary}[theorem]{Corollary}
\newtheorem{atheorem}{Theorem}[section]
\newtheorem{adefinition}[atheorem]{Definition}
\newtheorem{aproposition}[atheorem]{Proposition}
\newtheorem{alemma}[atheorem]{Lemma}
\newtheorem{acorollary}[atheorem]{Corollary}
\theoremstyle{definition}
\newtheorem{example}[theorem]{Example}

\renewcommand{\ket}[1]{|#1\rangle}
\renewcommand{\bra}[1]{\langle#1|}
\newcommand{\1}{\mathbbm{1}}
\DeclareMathOperator{\Span}{Span}
\DeclareMathOperator{\diag}{diag}

\usepackage{tikz}
\usetikzlibrary{decorations.pathreplacing,calligraphy,decorations.markings}

\definecolor{tensorcolor}{rgb}{0.65,0.77,0.95}
\definecolor{btensorcolor}{rgb}{0.65,0.50,0.69}
\definecolor{whitetensorcolor}{rgb}{0.93,0.93,0.93}

\newcommand{\gate}[2]{
    \begin{scope}[shift={(#1)}]
        \draw[ thick, fill=whitetensorcolor, rounded corners=2pt] (-\doubledx/2-0.2,0.27) rectangle (\doubledx/2+0.2,-.27); 
	    \draw (0,0) node {\scriptsize #2};
    \end{scope}
        }

\newcommand{\GTensor}[5]{
	\begin{scope}[shift={(#1)}]
    \ifnum#5=0
		\draw[very thick] (-#2,0) -- (#2,0);
		\draw[very thick] (0,#2) -- (0,-#2);
    \fi
    \ifnum#5=-1
		\draw[very thick] (0,0) -- (#2,0);
		\draw[very thick] (0,#2) -- (0,-#2);
    \fi
    \ifnum#5=1
		\draw[very thick] (-#2,0) -- (0,0);
		\draw[very thick] (0,#2) -- (0,-#2);
    \fi
        \draw[ thick, fill=tensorcolor, rounded corners=2pt] (-#3,-#3) rectangle (#3,#3);
		\draw (0,0) node {\scriptsize #4};
	\end{scope}
}

\newcommand{\GDTensor}[5]{
	\begin{scope}[shift={(#1)}]
    \ifnum#5=0
		\draw[very thick] (-#2,0) -- (#2,0);
		\draw[very thick] (0,#2) -- (0,-#2);
    \fi
    \ifnum#5=-1
		\draw[very thick] (0,0) -- (#2,0);
		\draw[very thick] (0,#2) -- (0,-#2);
    \fi
    \ifnum#5=1
		\draw[very thick] (-#2,0) -- (0,0);
		\draw[very thick] (0,#2) -- (0,-#2);
    \fi
        \draw[ thick, fill=tensorcolor, rounded corners=2pt] (-#3,-#3) rectangle (#3,#3);
    \def\dx{#3/3};
	\draw [thick]  (-#3+\dx, \dx) -- (- \dx,#3-\dx);
	\draw [thick] (-#3+1.5*\dx,-#3+1.5*\dx) -- (#3-1.5*\dx,#3-1.5*\dx);
	\draw [thick]  ( \dx, -#3 + \dx) -- (#3 - \dx,-\dx);
	\draw (0,0) node {\scriptsize #4};
	\end{scope}
}

\newcommand{\ATensor}[3]{
    \GTensor{#1}{1}{.5}{#2}{#3};
}

\newcommand{\ADTensor}[3]{
    \GDTensor{#1}{1}{.5}{#2}{#3};
}

\newcommand{\DoubleATensor}[2]{
	\begin{scope}[shift={(#1)}]
        \ADTensor{(0,.8)}{}{#2};
        \ATensor{(0,-.8)}{}{#2};
	\end{scope}
}

\newcommand{\ETensor}[2]{
	\begin{scope}[shift={(#1)}]
        \GDTensor{(0,.8)}{1}{.5}{}{#2};
        \GTensor{(0,-.8)}{1}{.5}{}{#2};
    \def\dx{.75};
	\draw [very thick] (0,-1.8) to  [bend left=90] (-\dx,-1.8);
	\draw [very thick] (0,1.8) to  [bend right=90] (-\dx,1.8);
	\draw [very thick] (-\dx,1.8) to  (-\dx,-1.8);
	\end{scope}
}

\newcommand{\STensor}[3]{
	\begin{scope}[shift={(#1)}]
		\draw [very thick] (0,3) -- (0,-0.8);
        \GDTensor{(0,.8)}{1}{.5}{}{#3};
        \GTensor{(0,-.8)}{1}{.5}{}{#3};
		\filldraw[color=black, fill=whitetensorcolor, thick] (0, 2.2) circle (\stradius);
      \def\dx{.8};
	\draw [very thick] (0,-1.8) to  [bend left=90] (-\dx,-1.8);
	\draw [very thick] (0,3) to  [bend right=90] (-\dx,3);
	\draw [very thick] (-\dx,3) to  (-\dx,-1.8);
	\draw (0,2.2) node {#2};
	\end{scope}
}

\newcommand{\bTensor}[2]{
	\begin{scope}[shift={(#1)}]
	    \draw [very thick] (-1,0) to  (1,0);
		\filldraw[color=black, fill=btensorcolor, thick] (0,0) circle (\stradius);
	\draw (0,0) node {#2};
	\end{scope}
}

\newcommand{\SingleTrLeft}[1]{
	\begin{scope}[shift={(#1)}]
      \draw [very thick] (0,0) to (\doubledx-0.8,0);
	   \draw [very thick] (0,0) to  [bend left=90] (0,0.8);
	   \draw [very thick, dotted] (0,0.8) to  (1,0.8);
	\end{scope}
}

\newcommand{\SingleDots}[2]{
	\begin{scope}[shift={(#1)}]
      \draw [very thick, dotted] (-#2/2,0) to (#2/2,0);
	\end{scope}
}

\newcommand{\DoubleDots}[2]{
	\begin{scope}[shift={(#1)}]
      \SingleDots{0,0.8}{#2};
      \SingleDots{0,-0.8}{#2};
	\end{scope}
}

\newcommand{\SingleTrRight}[1]{
	\begin{scope}[shift={(#1)}, xscale=-1]
	   \SingleTrLeft{(0,0)};
	\end{scope}
}

\newcommand{\IdentityTensor}[3]{
	\begin{scope}[shift={(#1)}]
      \draw [very thick] (0,-1) to (0,1);
    \ifnum#3=1
        \draw [very thick] (0,-1) to (0,1);
	    \filldraw[color=black, fill=whitetensorcolor, thick] (0,0) circle (\stradius);
	    \draw (0,0) node {#2};
    \fi
	\end{scope}
}

\newcommand{\DoubleIdentityTensor}[3]{
	\begin{scope}[shift={(#1)}]
        \draw [very thick] (0,-1.8) to (0,1.8);
    \ifnum#3=1
        \draw [very thick] (0,-1.8) to (0,1.8);
	    \filldraw[color=black, fill=whitetensorcolor, thick] (0,0) circle (\stradius);
	    \draw (0,0) node {#2};
    \fi
	\end{scope}
}

\newcommand{\DoubleLongLine}[1]{
	\begin{scope}[shift={(#1)}]
        \def\dx{.75};
	    \draw [very thick] (0,-1.8) to  [bend left=90] (0-\dx,-1.8);
	    \draw [very thick] (0,3.2) to  [bend right=90] (0-\dx,3.2);
	    \draw [very thick] (-\dx,3.2) to  (-\dx,-1.8);
	    \draw [very thick] (0,3.2) to  (0,2.8);
    \end{scope}
}

\newcommand{\SideIdentityTensor}[4]{
	\begin{scope}[shift={(#1)}]
    \ifnum#4=-1
	   \draw [very thick] (\doubledx-1,0.8) to  [bend right=90] (\doubledx-1,-0.8);
    \fi
    \ifnum#4=-2
	   \draw [very thick] (\doubledx-1,0.8) to  [bend right=90] (\doubledx-1,-0.8);
      \draw [very thick] (\doubledx-1,0.8) -- (\doubledx-0.5,0.8);
      \draw [very thick] (\doubledx-1,-0.8) -- (\doubledx-0.5,-0.8);
    \fi
    \ifnum#4=-3
	   \draw [very thick] (\doubledx-1,0.8) to  [bend right=90] (\doubledx-1,-0.8);
      \draw [very thick] (\doubledx-1,0.8) -- (\doubledx-0.5,0.8);
      \draw [very thick] (\doubledx-1,-0.8) -- (\doubledx-0.5,-0.8);
	\filldraw[color=black, fill=whitetensorcolor, thick] (\doubledx-1.4,0) circle (#3);
	\draw (\doubledx-1.4,0) node {#2};
    \fi
    \ifnum#4=1
	   \draw [very thick] (-\doubledx+1,0.8) to  [bend left=90] (-\doubledx+1,-0.8);
    \fi
    \ifnum#4=2
	   \draw [very thick] (-\doubledx+1,0.8) to  [bend left=90] (-\doubledx+1,-0.8);
      \draw [very thick] (-\doubledx+1,0.8) -- (-\doubledx+0.5,0.8);
      \draw [very thick] (-\doubledx+1,-0.8) -- (-\doubledx+0.5,-0.8);
    \fi
    \ifnum#4=3
	   \draw [very thick] (-\doubledx+1,0.8) to  [bend left=90] (-\doubledx+1,-0.8);
      \draw [very thick] (-\doubledx+1,0.8) -- (-\doubledx+0.5,0.8);
      \draw [very thick] (-\doubledx+1,-0.8) -- (-\doubledx+0.5,-0.8);
	\filldraw[color=black, fill=whitetensorcolor, thick] (-\doubledx+1.4,0) circle (#3);
	\draw (-\doubledx+1.4,0) node {#2};
    \fi
\end{scope}
}

\newcommand\doubledx{1.6}
\newcommand\singledx{1.8}
\newcommand\identitydx{1}
\newcommand\stradius{0.6}

\newcommand\subsetsim{\mathrel{%
  \ooalign{\raise0.2ex\hbox{$\subset$}\cr\hidewidth\raise-0.8ex\hbox{\scalebox{0.9}{$\sim$}}\hidewidth\cr}}}

\begin{document}

\title{Matrix-product unitaries: Beyond quantum cellular automata}

\author{Georgios Styliaris}
\affiliation{Max Planck Institute of Quantum Optics, Hans-Kopfermann-Str. 1, Garching 85748, Germany}
\affiliation{Munich Center for Quantum Science and Technology (MCQST), Schellingstr. 4, 80799 M{\"{u}}nchen, Germany}
\orcid{0000-0002-6809-8505}

\author{Rahul Trivedi}
\affiliation{Max Planck Institute of Quantum Optics, Hans-Kopfermann-Str. 1, Garching 85748, Germany}
\affiliation{Munich Center for Quantum Science and Technology (MCQST), Schellingstr. 4, 80799 M{\"{u}}nchen, Germany}
\affiliation{Electrical and Computer Engineering, University of Washington, Seattle, Washington 98195, USA}
\orcid{0000-0002-5621-7255}

\author{David P\'erez-Garc\'ia}
\affiliation{Departamento de An\'alisis Matem\'atico, Universidad Complutense de Madrid, 28040 Madrid, Spain}
\affiliation{Instituto de Ciencias Matem\'aticas (CSIC-UAM-UC3M-UCM), 28049 Madrid, Spain}
\orcid{0000-0003-2990-791X}

\author{J.~Ignacio Cirac}
\affiliation{Max Planck Institute of Quantum Optics, Hans-Kopfermann-Str. 1, Garching 85748, Germany}
\affiliation{Munich Center for Quantum Science and Technology (MCQST), Schellingstr. 4, 80799 M{\"{u}}nchen, Germany}
\orcid{0000-0003-3359-1743}

\maketitle

\begin{abstract}
Matrix-product unitaries (MPU) are 1D tensor networks describing time evolution and unitary symmetries of quantum systems, while their action on states by construction preserves the entanglement area law. MPU which are formed by a single repeated tensor are known to coincide with 1D quantum cellular automata (QCA), i.e., unitaries with an exact light cone. However, this correspondence breaks down for MPU with open boundary conditions, even if the resulting operator is translation-invariant. Such unitaries can turn short- to long-range correlations and thus alter the underlying phase of matter. Here we make the first steps towards a theory of MPU with uniform bulk but arbitrary boundary. In particular, we study the structure of a subclass with a direct-sum form which maximally violates the QCA property. We also consider the general case of MPU formed by site-dependent (nonuniform) tensors and show a correspondence between MPU and locally maximally entanglable states.
\end{abstract}

\section{Introduction}

Tensor networks have been successful in describing a wide variety of phenomena of many-body quantum systems with local interactions~\cite{orus2019tensor}. One of the reasons is that they can capture the complex structure of the underlying correlations while allowing for a compact description of relevant states and operations over the vast many-body Hilbert space. At the same time, the tensor-network framework naturally allows the utilization of tools from quantum information theory, which underlie many analytical results~\cite{cirac2021matrix}.

The paradigmatic tensor-network family in 1D is matrix-product states (MPS)~\cite{fannes1992finitely}. This family captures states satisfying an entanglement area law~\cite{verstraete2006matrix} which faithfully approximate low-energy states of local Hamiltonians~\cite{hastings2007area}. Many key theoretical results for MPS are obtained by examining the freedom in the representation of a given physical state in terms of different, but equivalent, sets of tensors~\cite{fannes1992finitely,vidal2003efficient,perez_garcia2007matrix}. By imposing a convenient representation, it follows that every MPS can be physically implemented as a quantum circuit with a linear (in the system size) number of gates~\cite{schon2005sequential}. This, thus, imposes a tight upper bound on their circuit complexity. Moreover, for uniform MPS (composed of a single repeated tensor), the nature of their correlation (short- versus long-range and the associated correlation length) can be directly extracted from a single tensor after imposing a canonical form~\cite{cirac2017matrix2}. Understanding the canonical form has also made possible the complete classification of topological phases in the framework of MPS, as well as their symmetry-enriched counterpart~\cite{chen2011classification,schuch2011classifying}. In both cases, the corresponding phase can be explicitly identified at the tensor level. An essential physical insight for both classifications is a renormalization-group transformation for MPS~\cite{verstraete2005renormalization,cirac2017matrix2}, which reveals the long-range correlation structure and also allows for improved preparation schemes~\cite{malz2024preparation}. 

The unitary counterpart of 1D MPS are matrix-product unitaries (MPU)~\cite{cirac2017matrix1,sahinoglu2018matrix,piroli2021fermionic}. The key feature of this family is that it preserves the entanglement area law of the underlying states. This makes MPU both a physically relevant and a broad class of unitaries which encompasses, for instance, shallow quantum circuits and short-time evolutions of local Hamiltonians. It is therefore desirable to study the structure and physical properties of MPU, a task which can be approached via tools from tensor networks and quantum information theory.

Many properties of MPU are well-understood for uniform tensor networks, i.e., those composed of a single repeated tensor and periodic boundary. In that case MPU coincide with translation-invariant (TI) quantum cellular automata (QCA)~\cite{cirac2017matrix1}, i.e., evolutions with an exact light cone~\cite{farrelly2020review}. This has many direct implications for their correlation structure and circuit complexity. First, it is known that all QCA can be physically implemented as finite-depth local quantum circuits with the aid of ancillas~\cite{arrighi2011unitarity}. Second, it implies that uniform MPU cannot convert short- to long- range correlations, thus they preserve the underlying phase of matter~\cite{chen2011classification,schuch2011classifying}. Third, QCA in 1D have been fully classified~\cite{schumacher2004reversible}, and an elegant topological theory has been developed~\cite{gross2012index}. Thus these results automatically apply to uniform MPU, completely specifying their complexity and the short-ranged nature of their correlations. Moreover, a canonical form for MPU was obtained in Ref.~\cite{cirac2017matrix1}. This form permits to explicitly obtain the QCA topological index on the tensor level~\cite{cirac2017matrix1,gong2021topological} and has also made possible a symmetry classification~\cite{gong2020classification}.

The correspondence between QCA and MPU, however, fails to hold beyond uniform tensor networks, i.e., those with open boundary or, more generally, site-dependent tensors. Thus for this general class of MPU (i) a characterization of the tensor structure and correlations, (ii) their complexity and physical implementation, and (iii) a topological classification, are all missing. Here we make some first steps in exploring this class of unitary tensor networks. First, we show that there exist physically relevant MPU which are TI for all system sizes but violate the QCA property. Such MPU break the light cone and can thus convert short- to long-range correlations, altering the underlying phase of matter while still preserving the entanglement scaling (i.e., map area law states to area law states). The existence of TI MPU which are not QCA is possible because, for a constant bond dimension, such MPU can only be expressed with open boundary conditions, but not periodic. Motivated by this, we study MPU composed of a repeated bulk tensor and a fixed boundary. We identify and study the structure of a subclass of these operators, which we call semi-simple MPU, that we show is either a trivial product unitary, or necessarily violates the QCA property.

We also consider the general problem of MPU composed of site-dependent tensors. Utilizing the MPS canonical form, we observe that every nonuniform MPU can be expressed as a sequence of isometries of increasing size. We also show a correspondence between the Choi-Jamio{\l}kowski  state~\cite{nielsen2002quantum} of a MPU and the so-called locally maximally entanglable (LME) states, a class previously introduced in multipartite entanglement theory~\cite{kruszynska2009local}.

The paper is structured as follows. We first establish some preliminaries in \cref{sec:preliminaries} and then, in \cref{sec:motivating_example}, introduce a simple example of a TI MPU which is not a QCA. We then study general nonuniform MPU in \cref{sec:inhomo_structure} and MPU with uniform bulk and open boundary in \cref{sec:homo}. The connection between LME states and MPU appears in \cref{sec:lme}. We conclude in \cref{sec:outlook}.

\section{Preliminaries and notation} \label{sec:preliminaries}

We extensively use graphical notation, so we first establish some conventions. We will often be concerned with rank-4 tensors
\begin{align}
    \left(A_k^{ij}\right)_{mn} =
        \begin{array}{c}
        \begin{tikzpicture}[scale=0.6,baseline={([yshift=-0.65ex] current bounding box.center)}]
            \ATensor{0,0}{\small $A_k$}{0}
		\draw (-1.4,0) node {$m$};
		\draw (1.4,0) node {$n$};
		\draw (0,1.4) node {$i$};
		\draw (0,-1.4) node {$j$};
        \end{tikzpicture}
        \end{array}
        \in \mathbb C
\end{align}
where $k$ labels the tensor, $i,j = 1,\dots,d_{\rm{out/in}}$ correspond to the output/input physical space and $m,n = 1,\dots, D_{k/k-1}$ to the auxiliary space. Here $D_k$ ($D_{k-1}$) is the left (right) bond dimension of $A_k$ and thus $A_k^{ij} \in \mathbb M_{D_k,D_{k-1}}$ is interpreted as a matrix.

Given a sequence of tensor $A_1,\dots,A_N$ with compatible bond dimension we will be interested in the matrix-product operator (MPO)
\begin{align} \label{eq:unitary_def}
 O &= \sum_{i,j} \Tr\left( b A_N^{i_N j_N} \dots  A_1^{i_1 j_1} \right) \ket{i_N \dots i_1} \! \bra{j_N \dots j_1} ,
\end{align}
where $b \in \mathbb M_{D_0,D_N}$ is the boundary condition. The case $b = \ket{r} \bra{l}$ corresponds to open boundary while $b = \1$ (if $D_0 = D_N$) to periodic. Graphically
\begin{align}
O = 
    \begin{array}{c}
        \begin{tikzpicture}[scale=.5,baseline={([yshift=-0.75ex] current bounding box.center)}]
		      \foreach \x in {0,...,0}{
                \SingleTrRight{(0,0)}
        }
		      \foreach \x in {1,...,1}{
                \GTensor{(-\singledx*\x,0)}{1}{.6}{\small $A_\x$}{0}
        }
		      \foreach \x in {2,...,2}{
                \GTensor{(-\singledx*\x,0)}{1}{.6}{\small $A_\x$}{0}
        }
		      \foreach \x in {3,...,3}{
                \SingleDots{-\singledx*\x,0}{\singledx/2}
        }
		      \foreach \x in {4,...,4}{
                \GTensor{(-\singledx*\x,0)}{1}{.6}{\small $A_N$}{0}
        }
		      \foreach \x in {5,...,5}{
                \bTensor{-\singledx*\x,0}{b}
        }
		      \foreach \x in {6,...,6}{
                \SingleTrLeft{(-\singledx*\x+.3,0)}
        }
        \end{tikzpicture}
        \end{array}
        \,.
\end{align}
We call a MPO uniform if its tensors $A_k$ are equal and it has periodic boundary, and a matrix-product unitary (MPU) if $O^\dagger O = \1$. 

When interpreting rank-2 tensors in the graphical notation as operators, unless otherwise stated, we follow the convention
\begin{align}
    \begin{array}{c}
        \begin{tikzpicture}[scale=.6,baseline={([yshift=-0.75ex] current bounding box.center)}]
		\draw[very thick] (0,1) -- (0,-1);
        \draw[ thick, fill=tensorcolor, rounded corners=2pt] (-.5,-.5) rectangle (.5,.5);
    \end{tikzpicture}
    \end{array}
    = \sum_{ij}     \begin{array}{c}
        \begin{tikzpicture}[scale=0.6,baseline={([yshift=-0.75ex] current bounding box.center)}]
		\draw[very thick] (0,1) -- (0,-1);
        \draw[ thick, fill=tensorcolor, rounded corners=2pt] (-.5,-.5) rectangle (.5,.5);
		\draw (0,1.5) node {$i$};
		\draw (0,-1.5) node {$j$};
    \end{tikzpicture}
    \end{array}{\ket{i}\bra{j}}
    \;, \quad
    \begin{array}{c}
        \begin{tikzpicture}[scale=0.6,baseline={([yshift=-0.75ex] current bounding box.center)}]
		\draw[very thick] (1,0) -- (-1,0);
        \draw[ thick, fill=tensorcolor, rounded corners=2pt] (-.5,-.5) rectangle (.5,.5);
    \end{tikzpicture}
    \end{array}
    = \sum_{mn}     \begin{array}{c}
        \begin{tikzpicture}[scale=0.6,baseline={([yshift=-0.75ex] current bounding box.center)}]
		\draw[very thick] (1,0) -- (-1,0);
        \draw[ thick, fill=tensorcolor, rounded corners=2pt] (-.5,-.5) rectangle (.5,.5);
		\draw (1.5,0) node {$n$};
		\draw (-1.5,0) node {$m$};
    \end{tikzpicture}
    \end{array}{\ket{m}\bra{n}} \,,
\end{align}
i.e., we assume as input the bottom/right and as output the top/left. In all other situations, we will explicitly use arrows to indicate the input and output. We will also alternate without explicit distinction between operators
\begin{subequations}
\begin{align}
        X = \sum_{ij} X_{ij} \ket{i}\bra{j} = 
        \begin{array}{c}
        \begin{tikzpicture}[scale=0.52]
            \IdentityTensor{0,0}{$X$}{1};
        \end{tikzpicture}
        \end{array}
\end{align}
and their vectorized counterpart
\begin{align}
        \ket{X} = \sum_{ij} X_{ij} \ket{i}\ket{j} =
        \begin{array}{c}
        \begin{tikzpicture}[scale=0.52]
            \SideIdentityTensor{(0,0)}{$X$}{\stradius}{3}
        \end{tikzpicture}
        \end{array}
        \,.
\end{align}
\end{subequations}
We use shading to denote
\begin{align}
    \begin{array}{c}
        \begin{tikzpicture}[scale=0.6,baseline={([yshift=-0.75ex] current bounding box.center)}]
        \GDTensor{(0,0)}{1}{.5}{}{0};
		\draw (-1.5,0) node {$m$};
		\draw (1.5,0) node {$n$};
		\draw (0,1.5) node {$j$};
		\draw (0,-1.5) node {$i$};
    \end{tikzpicture}
    = \Bigg(
        \begin{tikzpicture}[scale=0.6,baseline={([yshift=-0.75ex] current bounding box.center)}]
            \GTensor{(0,0)}{1}{.5}{}{0};
		\draw (-1.5,0) node {$m$};
		\draw (1.5,0) node {$n$};
		\draw (0,1.5) node {$i$};
		\draw (0,-1.5) node {$j$};
    \end{tikzpicture}
    \Bigg)^* \,.
    \end{array}
\end{align}
Later, we will often have contractions of the form
\begin{align}
            \begin{array}{c}
        \begin{tikzpicture}[scale=0.6]
		      \foreach \x in {1,...,1}{
                \DoubleATensor{(-\doubledx*\x,0)}{1};
                \draw(-\x*\doubledx,-0.8)  node {\small $A_1$};
        }
		      \foreach \x in {2,...,2}{
                \DoubleATensor{(-\doubledx*\x,0)}{0};
                \draw(-\x*\doubledx,-0.8)  node {\small $A_2$};
        }
		      \foreach \x in {3,...,3}{
                \DoubleDots{-\doubledx*\x,0}{\doubledx/2};
        }
		      \foreach \x in {4,...,4}{
                \DoubleATensor{(-\doubledx*\x,0)}{-1};
                \draw(-\x*\doubledx,-0.8)  node {\small $A_N$};
        }
        \end{tikzpicture}
                \end{array}
        \;.
\end{align}
{In such cases, each shaded tensor corresponds to its unshaded partner directly below it.


%

\section{Motivating example} \label{sec:motivating_example}

Quantum cellular automata (QCA) are unitaries that capture the notion of a causal evolution~\cite{farrelly2020review}.
Formally, $U$ is a 1D QCA if, for any local observable $O_{i}$ with support over any single site $i$, the evolved operator $U^\dagger O_i U$ is supported at most over $i$ and its nearest neighbors\footnote{Note that this property also implies that the analogous restriction holds for observables $O_A$ with support over an arbitrary (multi-site) region $A$. More generally, the definition of QCA can include the case where the evolved operator is supported in a ball of finite radius $R$ around site $i$. The general case reduces to the nearest neighbor case by renormalization, i.e., blocking sites together.}.
Therefore QCA are quantum evolutions with an exact light cone; $t$ time steps cannot create correlations in any pair of regions separated by $r > 2 t$ sites.

Although QCA can be identified with tensor network unitaries in any spatial dimensions~\cite{piroli2020quantum}, in 1D the connection is especially explicit.
Consider MPU which are formed (for any $N>1$) by repetitions of a single tensor $A$ over $N$ sites and periodic-boundary conditions. Note that the resulting operators are translation-invariant (TI) by construction. In Ref.~\cite{cirac2017matrix1} it was shown that, after blocking together a constant number of sites, this class of MPU exactly coincides with TI QCA. That is, in 1D the light-cone property necessarily emerges by imposing unitarity over the matrix-product operator structure.

This result characterizes uniform MPU with periodic boundary. This is since QCA in 1D can only be formed by finite-depth circuits and the shift operation~\cite{schumacher2004reversible,farrelly2020review}. Nonetheless, consider the following situation.
\begin{example}[Multi-control $Z$-gate]\label{ex:multi_control_z}
Define
\begin{align}\label{eq:controlZ}
    U_{Z} = \1^{\otimes N} - 2 (\ket{1}\bra{1})^{\otimes N}
\end{align}
which is a control-$Z$ unitary with $N-1$ control qubits and a single target.
%
%
It is direct to verify that $U_{Z}$ has the following properties:
\begin{enumerate}[(i)]
    \item Permutation invariance, which also implies translation invariance.
    \item Operator Schmidt rank $D = 2$ across every bipartition.
    \item It is not a QCA since there exist control-qubits arbitrarily far from the target qubit, i.e., the light-cone property is violated at all distances.
\end{enumerate}
\end{example}

At first glance, this simple example appears to be in tension with the previous identification between MPU and QCA. The resolution lies in the fact that a $D=O(1)$ MPU representation of $U_{Z}$ has to necessarily involve open boundary conditions, although the unitary itself is TI. For instance, a $D=2$ representation is
    \begin{subequations} \label{eq:tensors_cz}
        \begin{align}
            A^{11} = \1 \;, \quad  A^{00} = \ket{0}\bra{0} \;, \quad A^{01} = A^{10} = 0 \;,
        \end{align}
with boundary conditions
    \begin{align}
            \bra{l} = \bra{0}+ \bra{1} \;, \quad \ket{r} = \ket{0} - 2\ket{1}  \,.
        \end{align}
    \end{subequations}
%
%
This situation is reminiscent of the $W$ state~\cite{dur2000three} in the context of MPS\footnote{The $W$ state is TI and has Schmidt rank equal to 2 across every bipartition. Although it can be written as a $D=2$ MPS with open boundary conditions and a single bulk tensor, it does not admit a representation in terms of a single tensor and periodic boundary unless $D$ scales with the system size~\cite{perez_garcia2007matrix,michalek2019quantum}.}.

It would thus be desirable to have a characterization of MPU beyond the paradigm of a uniform tensor with periodic boundary. Here we make the initial steps in this direction. We begin with the general case of nonuniform MPU (i.e., consisting of site-dependent tensors) and formulate unitarity as a condition over the auxiliary space. We then move to MPU that have uniform bulk but allow for a boundary operator. We give examples and systematically examine the structure of a subclass, which we call semi-simple MPU. This class generalizes \cref{ex:multi_control_z} and includes families of TI MPU which violate the QCA property. Finally we establish a connection with locally maximally entanglable (LME) states~\cite{kruszynska2009local}.

\section{Nonuniform MPU} \label{sec:inhomo_structure}

\subsection{Structure}

The aim here is to characterize the following property.
\begin{definition}[Nonuniform MPU]
    A sequence of tensors $A_{1}, \dots, A_{N}$ generates a MPU if
\begin{align}
 U = 
        \begin{array}{c}
        \begin{tikzpicture}[scale=0.5]
		      \foreach \x in {1,...,1}{
                \GTensor{(-\singledx*\x,0)}{1}{.6}{\small $A_\x$}{1}
        }
		      \foreach \x in {2,...,2}{
                \GTensor{(-\singledx*\x,0)}{1}{.6}{\small $A_\x$}{0}
        }
		      \foreach \x in {3,...,3}{
                \SingleDots{-\singledx*\x,0}{\singledx/2}
        }
		      \foreach \x in {4,...,4}{
                \GTensor{(-\singledx*\x,0)}{1}{.6}{\small $A_N$}{-1}
        }
        \end{tikzpicture}
                \end{array}
\end{align}
is unitary.
\end{definition}

Unless otherwise stated, we assume open boundary conditions\footnote{Note that periodic ones can be included as a special case at the expense of increasing the bond dimension $D_k \mapsto D_0 D_k$. Analogously for arbitrary $b$.} and that all tensors have equal input-output physical dimension $d$.
Although we will often only explicitly write relations for tensors in the bulk of the chain, all expressions should also be understood for the tensors in the boundary by substituting $D_0 = D_N = 1$.
Finally, we always implicitly assume $D \equiv \max D_k = O(1)$ (i.e., size-independent).


Our starting point is to formulate unitarity
\begin{align} \label{eq:unitarity}
 U^\dagger U = 
        \begin{array}{c}
        \begin{tikzpicture}[scale=0.6]
		      \foreach \x in {1,...,1}{
                \DoubleATensor{(-\doubledx*\x,0)}{1}
                \draw (-\doubledx*\x,-0.8) node {\small $A_1$};
        }
		      \foreach \x in {2,...,2}{
                \DoubleATensor{(-\doubledx*\x,0)}{0}
                \draw (-\doubledx*\x,-0.8) node {\small $A_2$};
        }
		      \foreach \x in {3,...,3}{
                \DoubleDots{-\doubledx*\x,0}{\doubledx/2}
        }
		      \foreach \x in {4,...,4}{
                \DoubleATensor{(-\doubledx*\x,0)}{-1}
                \draw (-\doubledx*\x,-0.8) node {\small $A_N$};
        }
        \end{tikzpicture}
                \end{array}
         = 
                 \begin{array}{c}
        \begin{tikzpicture}[scale=0.6]
		      \foreach \x in {1,...,2}{
                \DoubleIdentityTensor{(-\identitydx*\x,0)}{}{0};
        }
		      \foreach \x in {3,...,3}{
                \DoubleDots{-\identitydx*\x,0}{\identitydx/2};
        }
		      \foreach \x in {4,...,4}{
                \DoubleIdentityTensor{(-\identitydx*\x,0)}{}{0};
        }
        \end{tikzpicture}
        \end{array}
\end{align}
as a condition over individual tensors. For this, we decompose
\begin{align} \label{eq:local_decomposition}
        \begin{array}{c}
        \begin{tikzpicture}[scale=0.6]
            \GDTensor{(0,0.8)}{1}{.5}{}{0};
            \ATensor{(0,-0.8)}{\small $A_k$}{0};
        \end{tikzpicture}
        \end{array}
    = E^{(k)} \otimes \1 + \sum_{j=1}^{d^2-1} S_j^{(k)} \otimes \sigma_j \;,
\end{align}
where the first part of the tensor product acts on the auxiliary space and the second on the physical.
Here $\sigma_j$ are orthonormal Hermitian traceless operators which, together with $\sigma_0 \equiv \1$, form a basis.
Thus
\begin{align}
    S_0^{(k)} \equiv E^{(k)} = \frac{1}{d}
        \begin{array}{c}
        \begin{tikzpicture}[scale=0.6]
            \ETensor{0,0}{0};
            \ATensor{(0,-0.8)}{\small $A_k$}{0};
        \end{tikzpicture}
        \end{array}
    \;, \quad S_j^{(k)} =
        \begin{array}{c}
        \begin{tikzpicture}[scale=0.6]
            \STensor{0,0}{\small $\sigma_{j}$}{0};
            \ATensor{(0,-0.8)}{\small $A_k$}{0};
        \end{tikzpicture}
        \end{array}
        \;.
\end{align}
With these definitions, interpreting \cref{eq:unitarity} horizontally, we obtain the following set of conditions, which we use repeatedly throughout the paper.

\begin{lemma} \label{prop:unitarity_bond_S}
    A sequence of tensors $A_{1}, \dots, A_{N}$ generates a MPU if and only if
    \begin{subequations} \label{eq:unitarity_bond}
    \begin{numcases}{S^{(N)}_{j_N} \dots S^{(2)}_{j_2} S^{(1)}_{j_1} = }
           \label{eq:unitarity_bond_1}  1 & if  all $j_k=0$, \\
            \label{eq:unitarity_bond_2} 0 & otherwise.
    \end{numcases}
    \end{subequations}
\end{lemma}
\begin{proof}
    It follows by substituting the decomposition~\eqref{eq:local_decomposition} into \cref{eq:unitarity} and taking the inner product with all operator strings $\sigma _{j_N} \otimes \dots \otimes \sigma_{j_1}$ over the physical space.
\end{proof}

Seemingly, this amounts to an exponential in $N$ number of conditions, which is unsatisfactory. However, this is an artifact and can be reduced to recursively verifiable conditions. The idea is to replace individual products in Eq.~\eqref{eq:unitarity_bond_2} by their span. We thus define
\begin{subequations}
\begin{align}
    \rho^{(k)} &= E^{(k)}\cdot \dotso \cdot E^{(1)} \\
    \mathcal S^{(k)} &= \Span_j \{ S^{(k)}_{j_k} \dots S^{(2)}_{j_2} S^{(1)}_{j_1}: \; \exists j_i \ne 0\}
\end{align}
\end{subequations}
where the span is over the reals. It is direct to see that $\mathcal S^{(k+1)}$ can be built recursively from $\rho^{(k)}$ and $\mathcal S^{(k)}$, by keeping track of a basis for the latter. In these variables:
\begin{corollary} \label{prop:inhomo_recursive_unitarity}
    A sequence of tensors $A_{1}, \dots, A_{N}$ generates a MPU if and only if
    \begin{align}
        \rho^{(N)} = 1 \;, \quad \mathcal S^{(N)} = 0 \;. \label{eq:unitarity_span}
    \end{align}
\end{corollary}
\begin{proof}
    Clearly $\rho^{(k)} = E^{(k)} \dots E^{(1)}$ thus the condition $\rho^{(N)} = 1$ is identical to Eq.~\eqref{eq:unitarity_bond_1}, while
    $\mathcal S^{(N)} = 0$ is identical to Eq.~\eqref{eq:unitarity_bond_2}. 
\end{proof}
%


We now lift some of the redundancy in representing a MPU as a tensor network. This will lead to additional constraints on $\rho^{(k)}$ and $\mathcal S^{(k)}$. Our strategy will be to vectorize the MPU to a MPS and then impose the (left) MPS canonical form~\cite{perez_garcia2007matrix}. Since this is always possible and only amounts to a preferred representation of the same global unitary, the following canonical form can be assumed without loss of generality.
%
\begin{definition}[Nonuniform canonical form]
A MPU is in canonical form if all its tensors satisfy
\begin{align} \label{eq:gauge_nonuniform}
        \frac{1}{d}
            \begin{array}{c}
            \begin{tikzpicture}[scale=0.5]
    		      \foreach \x in {1,...,1}{
                  \ETensor{-\doubledx*\x,0}{0};
                 }
    		      \foreach \x in {2,...,2}{
                  \SideIdentityTensor{-\doubledx*\x,0}{}{}{-1};
                 }
            \end{tikzpicture}
            \end{array}
        =
            \begin{array}{c}
            \begin{tikzpicture}[scale=.5]
                \SideIdentityTensor{0,0}{}{}{-2};
            \end{tikzpicture}
            \end{array}
        \;.
\end{align}
\noindent For tensors on the boundary, the same condition applies with the left/right auxiliary space interpreted as trivial.
\end{definition}
The procedure to bring a MPU to canonical form is identical to the one of MPS~\cite{perez_garcia2007matrix} after vectorizing the former. Explicitly, this can be achieved by iterating a singular-value decomposition $T_k = X_k \Sigma_k Y_k^\dagger$ ($X_k,Y_k$ isometries) with each operator $T_k$ arising from tensors with input/output as
\begin{align}
    \begin{array}{c}
        \begin{tikzpicture}[scale=0.6,baseline={([yshift=-0.65ex] current bounding box.center)}]
            \ATensor{0,0}{\small $T_k$}{0}
            \draw[thick] (.8,-.15) -- (.7,0) -- (0.8,0.15);
            \draw[thick] (-.7,-.15) -- (-.8,0) -- (-0.7,0.15);
            \draw[thick] (-0.15,0.7) -- (0,0.8) -- (0.15,0.7);
            \draw[thick] (-0.15,-0.7) -- (0,-0.8) -- (0.15,-0.7);
        \end{tikzpicture}
        =
         \begin{tikzpicture}[scale=0.6,baseline={([yshift=-.7ex] current bounding box.center)}]
            \draw[very thick] (-1,0) -- (4,0);
            \ATensor{0,0}{\small $X_k$}{0}
            \draw[thick] (.8,-.15) -- (.7,0) -- (0.8,0.15);
            \draw[thick] (-.7,-.15) -- (-.8,0) -- (-0.7,0.15);
            \draw[thick] (-0.15,0.7) -- (0,0.8) -- (0.15,0.7);
            \draw[thick] (-0.15,-0.7) -- (0,-0.8) -- (0.15,-0.7);
            \filldraw[color=black, fill=whitetensorcolor, thick] (1.5,0) circle (.5);
            \draw (1.5,0) node {\small $\Sigma_k$};
            \draw[thick] (2.3,-.15) -- (2.2,0) -- (2.3,0.15);
        \draw[ thick, fill=whitetensorcolor, rounded corners=2pt] (2.5,-.5) rectangle (3.5,0.5);
		\draw (3,0) node {\small $Y_k^\dagger$};
            \draw[thick] (3.8,-.15) -- (3.7,0) -- (3.8,0.15);
            \end{tikzpicture}
    \end{array}
\; .
\end{align}
\noindent The algorithm starts by setting $T_N = A_N$; the resulting isometry $X_N$ is taken to be the new tensor $A'_N = X_N$ which, by construction, satisfies \cref{eq:gauge_nonuniform} (up to a constant).
Iterating with $T_{k} = \Sigma_{k+1} Y_{k+1}^\dagger A_k$ and setting $A'_k = X_k$ results in a new MPU representation $A'_1,\dots,A'_N$.
In the last step, a scalar factor $\Tr U^\dagger U = d^N$ arises. Distributing the factor equally among the $A'_k$ results in the canonical form of \cref{eq:gauge_nonuniform}.
We will nevertheless consider a MPU in canonical form regardless of how this factor is distributed among the different tensors.

The interplay between canonical form and unitarity leads to the following conditions.

\begin{restatable}[]{proposition}{inhomocanonical}\label{prop:inhomocanonical}
    If a MPU is in canonical form, then for all $k =1,\dots, N$:
    \begin{enumerate}[(i)]
        \item The tensors $A_1,\dots,A_k$ satisfy
\begin{subequations}\label{subeq:inhomo_tp}
\begin{align} \label{eq:inhomo_tp}
            \begin{array}{c}
        \begin{tikzpicture}[scale=0.6]
		      \foreach \x in {1,...,1}{
                \DoubleATensor{(-\doubledx*\x,0)}{1};
                \draw(-\x*\doubledx,-0.8)  node {\small $A_1$};
        }
		      \foreach \x in {2,...,2}{
                \DoubleATensor{(-\doubledx*\x,0)}{0};
                \draw(-\x*\doubledx,-0.8)  node {\small $A_2$};
        }
		      \foreach \x in {3,...,3}{
                \DoubleDots{-\doubledx*\x,0}{\doubledx/2};
        }
		      \foreach \x in {4,...,4}{
                \DoubleATensor{(-\doubledx*\x,0)}{-1};
                \draw(-\x*\doubledx,-0.8)  node {\small $A_k$};
        }
		      \foreach \x in {5,...,5}{
                \SideIdentityTensor{(-\doubledx*\x,0)}{}{}{-2};
        }
        \end{tikzpicture}
                \end{array}
         = 
                 \begin{array}{c}
        \begin{tikzpicture}[scale=0.6]
		      \foreach \x in {1,...,2}{
                \DoubleIdentityTensor{(-\identitydx*\x,0)}{}{0};
        }
		      \foreach \x in {3,...,3}{
                \DoubleDots{-\identitydx*\x,0}{\identitydx/2};
        }
		      \foreach \x in {4,...,4}{
                \DoubleIdentityTensor{(-\identitydx*\x,0)}{}{0};
        }
        \end{tikzpicture}
        \end{array}
        \;,
\end{align}
which is equivalent to
\begin{align} \label{eq:inhom_unitarity_local}
        \begin{array}{c}
        \begin{tikzpicture}[scale=0.6]
            \foreach \x in {-1,...,-1}{
            \SideIdentityTensor{\x*\doubledx,0}{}{}{-1};
             }
		      \foreach \x in {0,...,0}{
            \DoubleATensor{-\x*\doubledx,0}{1};
            \draw (-\doubledx*\x,-0.8) node {\small $A_k$};
        }
		      \foreach \x in {1,...,1}{
	\begin{scope}[shift={(\x*\doubledx,0)}]
	   \draw [very thick] (-\doubledx+1,0.8) to  [bend left=90] (-\doubledx+1,-0.8);
      \draw [very thick] (-\doubledx+1,0.8) -- (-\doubledx+0.5,0.8);
      \draw [very thick] (-\doubledx+1,-0.8) -- (-\doubledx+0.5,-0.8);
	\filldraw[color=black, fill=whitetensorcolor, thick] (-\doubledx+1.9,0) ellipse (1.2 and 0.7);
	\draw (-\doubledx+1.95,0) node {\small $\rho^{{(k-1)}}$};
\end{scope}
        }
        \end{tikzpicture}
        \end{array}
    =
             \begin{array}{c}
            \begin{tikzpicture}[scale=0.6]
                \DoubleIdentityTensor{0,0}{}{0}
            \end{tikzpicture}
            \end{array}
    \;,\quad
            \begin{array}{c}
            \begin{tikzpicture}[scale=0.6]
    		      \foreach \x in {0,...,0}{
            \SideIdentityTensor{\x*\doubledx,0}{}{\stradius}{-2}
            }
    		      \foreach \x in {1,...,1}{
            \DoubleATensor{\x*\doubledx,0}{0}
                \draw (\doubledx*\x,-0.8) node {\small $A_k$};
            }
    		      \foreach \x in {2,...,2}{
            \SideIdentityTensor{\x*\doubledx,0}{\normalsize $X$}{\stradius}{3}
            }
            \end{tikzpicture}
            \end{array}
            = 0 \;, \; X \in \mathcal S^{(k-1)}\;.
\end{align}
\end{subequations}
        \item $\rho^{(k)}$ is a normalized quantum state and $\mathcal S^{(k)}$ is composed of traceless Hermitian operators.
    \end{enumerate}
\end{restatable}
\begin{proof}
$(i)$ The first equality follows by tracing out sites $k+1,\dots,N$ in \cref{eq:unitarity} and then simplifying from the left end using the gauge conditions. The second equation follows similarly by contracting with traceless operators $\sigma^{(k-1)}_{j_{k-1}}, \dots, \sigma^{(1)}_{j_1}$ on the right end of the chain.

$(ii)$ Consider the following linear map, mapping an operator over $k$ physical sites to an operator over the auxiliary space:
\begin{align}
    X \mapsto \mathcal E_k(X) =
    \begin{array}{c}
    \begin{tikzpicture}[scale=0.6]
        \draw[thick, fill=whitetensorcolor, rounded corners=2pt] (-3.7*\doubledx,2.8) rectangle (0.5*\doubledx,1.8);
	    \draw (-2*\doubledx,2.3) node {\small $X$};
		      \foreach \x in {0,...,0}{
                \DoubleLongLine{-\doubledx*\x,0}
                \DoubleATensor{(-\doubledx*\x,0)}{1}
                \draw (-\doubledx*\x,-0.8) node {\small $A_1$};
        }
		      \foreach \x in {1,...,1}{
                \DoubleLongLine{-\doubledx*\x,0}
                \DoubleATensor{(-\doubledx*\x,0)}{0}
                \draw (-\doubledx*\x,-0.8) node {\small $A_2$};
        }
		      \foreach \x in {2,...,2}{
                \DoubleDots{-\doubledx*\x,0}{\doubledx/2}
        }
		      \foreach \x in {3,...,3}{
                \DoubleLongLine{-\doubledx*\x,0}
                \DoubleATensor{(-\doubledx*\x,0)}{0}
                \draw (-\doubledx*\x,-0.8) node {\small $A_k$};
        }
    \end{tikzpicture}
    \end{array}
    \;.
\end{align}
By construction $\mathcal E_k$ preserves Hermiticity and $\Tr \mathcal E_k(X) = \Tr X$ due to \cref{eq:inhomo_tp}, i.e., it is also trace-preserving\footnote{According to our graphical convention of interpreting tensors as operators, $[\mathcal E_k(\cdot)]^T$ is completely-positive trace-preserving.}. The claim follows since $\rho^{(k)} = \mathcal E_{k-1} (\1^{\otimes k-1}/d^{k-1})$, $\rho^{(k)}$ is positive by construction, and $\mathcal S^{(k)} = \mathcal E_{k-1} (X)$ for $X$ Hermitian and traceless.
\end{proof}

A direct consequence is that:
\begin{corollary} \label{cor:sequence_isometries}
    Let $A_1,\dots,A_N$ be a MPU in canonical form and define
\begin{align} \label{eq:sequence_isometries}
 V_k = 
        \begin{array}{c}
        \begin{tikzpicture}[scale=0.5]
		      \foreach \x in {1,...,1}{
                \GTensor{(-\singledx*\x,0)}{1}{.6}{\small $A_\x$}{1};
        }
		      \foreach \x in {2,...,2}{
                \GTensor{(-\singledx*\x,0)}{1}{.6}{\small $A_\x$}{0};
        }
		      \foreach \x in {3,...,3}{
                \SingleDots{-\singledx*\x,0}{\singledx/2};
        }
		      \foreach \x in {4,...,4}{
                \GTensor{(-\singledx*\x,0)}{1}{.6}{\small $A_k$}{-0};
            \draw[very thick] (-\singledx*\x-1,0) -- (-\singledx*\x-1,1);
        }
        \end{tikzpicture}
                \end{array}
        \,.
\end{align}
Then $V_k^\dagger V_k = \1$, i.e., every MPU can be represented as a sequence of isometries $V_k$ $(k=1,\dots, N)$ with $U = V_N$.
\end{corollary}

This representation demonstrates how the canonical form of MPU strongly constrains their structure.
As we will see, however, this does not necessarily imply that every tensor $A_k$ is \textit{individually} an isometry when interpreted as a map ``diagonally" $A_k: \mathbb C^{d_{\rm in}} \otimes \mathbb C^{D_{k-1}} \to \mathbb C^{d_{\rm out}} \otimes \mathbb C^{D_k}$ with
$ A_k = 
    \begin{array}{c}
        \begin{tikzpicture}[scale=0.5,baseline={([yshift=-0.65ex] current bounding box.center)}]
            \ATensor{0,0}{\scriptsize $A_k$}{0}
            \draw[thick] (.8,-.15) -- (.7,0) -- (0.8,0.15);
            \draw[thick] (-.7,-.15) -- (-.8,0) -- (-0.7,0.15);
            \draw[thick] (-0.15,0.7) -- (0,0.8) -- (0.15,0.7);
            \draw[thick] (-0.15,-0.8) -- (0,-0.7) -- (0.15,-0.8);
        \end{tikzpicture}
    \end{array}
$. 
This observation will become important later when we discuss the sequential generation of MPU.


\subsection{When does a sequence of tensors generate a unitary?}

Here we provide necessary and sufficient conditions such that a given sequence of tensors $A_1,\dots,A_k$ can be extended to a MPU $A_1,\dots,A_N$ for some arbitrary (but finite) $N>k$. As we will see, this is not always possible, even if the given tensors are all in canonical form, which we assume without loss of generality.

A first observation is that \cref{prop:inhomocanonical} restricts the possible pairs of states $\rho^{(k)}$ and spans $\mathcal S^{(k)}$ that can emerge in a MPU. This is illustrated in: 

\begin{example} \label{ex:depolarizing}
Consider a MPU with $A_1,\dots, A_N$ in canonical form with each tensor having equal input-output physical dimension $d_{\rm {out}} = d_{\rm {in}}$. Then $\mathcal S^{(k)}$ cannot span the full space of traceless Hermitian operators for any $k < N$ unless $D_k = 1$, i.e., the MPU factorizes. This is because \cref{eq:inhom_unitarity_local} applied on the combined tensor $\widetilde{A} = A_{N} \cdot \dotso \cdot A_{k+1}$ gives
\begin{align}
            \begin{array}{c}
            \begin{tikzpicture}[scale=0.6]
    		      \foreach \x in {1,...,1}{
                  \DoubleATensor{-\doubledx*\x,0}{-1};
                 }
            \draw(-\doubledx,-0.82) node {\small $\widetilde{A}$};
            \end{tikzpicture}
            \end{array}
        =
            \begin{array}{c}
            \begin{tikzpicture}[scale=0.6]
                \DoubleIdentityTensor{(0,0)}{}{0};
                    \SideIdentityTensor{0.4,0}{}{}{-2};
            \end{tikzpicture}
            \end{array}      
            \,.
\end{align}
However, this is impossible unless the output physical dimension of $\widetilde{A}$ exceeds the input one.
\end{example}

In the rest of the section, we generalize this example to a complete set of conditions for $\rho^{(k)}$ and $\mathcal S^{(k)}$.

According to \cref{cor:sequence_isometries}, any valid tensor $A_{k+1}$ must extend the given isometry $V_{k}$ [cf.~\cref{eq:sequence_isometries}] to a new isometry $V_{k+1}$.
For instance, any $A_{k+1}$ isometric, i.e., satisfying 
\begin{align} \label{eq:local_isometry}
            \begin{array}{c}
            \begin{tikzpicture}[scale=0.7]
    		      \foreach \x in {1,...,1}{
                  \DoubleATensor{-\doubledx*\x,0}{0};
                \draw (-\doubledx*\x,-0.8) node {\scriptsize $A_{k+1}$};
                 }
    		      \foreach \x in {2,...,2}{
                  \SideIdentityTensor{-\doubledx*\x,0}{}{}{-1};
                 }
            \end{tikzpicture}
            \end{array}
        =
            \begin{array}{c}
            \begin{tikzpicture}[scale=0.7]
                \DoubleIdentityTensor{(0,0)}{}{0};
                    \SideIdentityTensor{0.4,0}{}{}{-2};
            \end{tikzpicture}
            \end{array}      
\end{align}
with $d_{{\rm out}} = d_{{\rm in}} = d$ and $D_{k+2} \ge D_{k+1}$ is a valid extension of the sequence.
However, the key point is that special care needs to be taken at the boundary $k=N$ since the resulting global operator $U = V_N$ must be unitary, and not only an isometry. This observation results in the following constraints, which are not only necessary but also sufficient.


%
\begin{restatable}[]{proposition}{channelmpu} \label{prop:growing}
    Consider a sequence of tensors $A_1,\dots, A_k$ in canonical form with equal input-output physical dimension. The sequence can be extended to a MPU $A_1,\dots, A_N$ ($N \ge k$) in canonical form if and only if there exists $d'\ge 1$ and a quantum channel $\mathcal T: \mathbb C^{D_k^2} \to \mathbb C^{d'^2}$ with Kraus rank $r_K = d'$ such that
    \begin{align} \label{eq:channel_extension}
        \mathcal T (\rho^{(k)}) = \1_{d'} \quad \text{and}\quad \mathcal T (X) = 0 \;\; \forall X \in \mathcal S^{(k)} \;.
    \end{align}
\end{restatable}
\begin{proof}
Suppose the sequence $A_1,\dots, A_k$ can be extended to $A_1,\dots,A_k,\dots,A_N$ forming a MPU in canonical form. For the completion $\widetilde{A}_L = A_{N}\dotso A_{k+1}$, due to the canonical form assumption, it holds that
\begin{align} \label{eq:app:left_tp}
    \frac{1}{d'}
        \begin{array}{c}
        \begin{tikzpicture}[scale=0.7]
		      \foreach \x in {2,...,2}{
                \ETensor{(-\doubledx*\x,0)}{0}
                \draw (-\doubledx*\x,-0.8) node {\scriptsize $A_{k+1}$};
        }
		      \foreach \x in {3,...,3}{
                \DoubleDots{-\doubledx*\x,0}{\doubledx/2}
        }
		      \foreach \x in {4,...,4}{
                \ETensor{(-\doubledx*\x,0)}{-1}
                \draw (-\doubledx*\x,-0.8) node {\small $A_N$};
        }
        \end{tikzpicture}
                \end{array}
         = 
            \begin{array}{c}
            \begin{tikzpicture}[scale=0.7]
                    \SideIdentityTensor{0,0}{}{}{-2};
            \end{tikzpicture}
            \end{array}
    \end{align}
where $d'$ is the physical dimension of $\widetilde{A}_L$. Note that $\widetilde{A}_L$ necessarily has equal input-output physical dimension because (globally) we have a MPU and the tensor $\widetilde{A}_R = A_k \dotso A_1$ has equal input-output dimension by assumption.

Because of \cref{eq:app:left_tp}, the map $X \mapsto \mathcal T(X)$ defined by
\begin{align}
   \mathcal T(X) =  \frac{1}{d'}
        \begin{array}{c}
        \begin{tikzpicture}[scale=0.7]
		      \foreach \x in {0,...,0}{
            \SideIdentityTensor{\x*\doubledx,0}{\normalsize $X$}{\stradius}{3};
        }
		      \foreach \x in {1,...,1}{
                \DoubleATensor{(-\doubledx*\x,0)}{0};
                \draw (-\doubledx*\x,-0.8) node {\scriptsize $A_{k+1}$};
        }
		      \foreach \x in {2,...,2}{
                \DoubleDots{-\doubledx*\x,0}{\doubledx/2};
        }
		      \foreach \x in {3,...,3}{
                \DoubleATensor{(-\doubledx*\x,0)}{-1};
                \draw (-\doubledx*\x,-0.8) node {\small $A_{N}$};
        }
        \end{tikzpicture}
                \end{array}
    \end{align}
is completely positive and trace-preserving, i.e., a quantum channel, with Kraus rank $r_K = d'$. This last property holds since $(i)$ $r_K \le d'$ cannot exceed the output physical dimension of $\widetilde {A}$, which here plays the role of the Kraus operator index, and $(ii)$ unitarity \cref{eq:unitarity}, together with the fact that both $\widetilde{A}_L,\widetilde{A}_R$ have equal input-output dimension implies $r_K \ge d'$. Finally, \cref{eq:channel_extension} follows by applying \cref{eq:inhom_unitarity_local} to $\widetilde {A}_L$.

Conversely, the existence of the quantum channel $\mathcal T$ with the given Kraus rank implies its decomposition over $d'$ Kraus operators, which define the tensor
\begin{align}
    K_i^\dagger = \frac{1}{\sqrt{d'}}
        \begin{array}{c}
        \begin{tikzpicture}[scale=.6,baseline={([yshift=-6.5ex] current bounding box.center)}]
            \ATensor{0,0}{\small $\widetilde A$}{-1}
	        \draw (0,1.4) node {\small $i$};
            \draw[thick] (.7,-.15) -- (.8,0) -- (0.7,0.15);
            \draw[thick] (-0.15,-0.8) -- (0,-0.7) -- (0.15,-0.8);
        \end{tikzpicture}
                \end{array}
            .
    \end{align}
Here the scalar factor is added for convenience. Then the sequence $A_1,\dots, A_k,\widetilde A$ is a MPU. This is because \cref{eq:inhom_unitarity_local} is satisfied for the tensor $\widetilde A$ (due to the assumption \cref{eq:channel_extension}) which implies the unitarity conditions of \cref{prop:inhomo_recursive_unitarity}.
\end{proof}

Notice that the nontrivial constraint here is the Kraus rank. For instance, in \cref{ex:depolarizing}, the unique corresponding channel $\mathcal T$ that satisfied the constrains of \cref{eq:inhom_unitarity_local} is the completely depolarizing. However, $r_K = d'^2$, thus failing the conditions of the Proposition.

%
%
%

%

\subsection{Can MPU be sequentially generated?} \label{subsec:inhomo_examples}

The sequential circuit has special importance in the theory of MPS. This is because every MPS can be generated sequentially with gates simultaneously acting on $\lceil \log_d D \rceil + 1$ consecutive qudits with a product initial state~\cite{schon2005sequential}. Here we investigate if an analogous result holds for MPU. For the case $D=d$, this corresponds to an MPU of the form:


\begin{example}[1-floor staircase] \label{ex:1_floor}
Consider the family of quantum circuits
\begin{align} \label{eq:1_floor_staicase}
                 \begin{array}{c}
        \begin{tikzpicture}[scale=.6]
		      \foreach \x in {1,...,5}{
      \draw [thick] (-\doubledx*\x,0) -- (-\doubledx*\x,5);
        }
		\foreach \x in {1,...,2}{
        \gate{(-\doubledx*\x-\doubledx*0.5,\doubledx/2*\x)}{\scriptsize $U_\x$};          
        }
		\fill[fill=white] (-3.1*\doubledx,2.8) rectangle (-2.9*\doubledx,2.2);
        \draw [very thick, dotted] (-2.8*\doubledx,2.2) to (-3.2*\doubledx,2.8);
        \begin{scope}[shift={(0,1)}]
		\foreach \x in {3,...,3}{
        \gate{(-\doubledx*\x-\doubledx*0.5,\doubledx/2*\x)}{\scriptsize $U_{N-1}$};          
        }
		\foreach \x in {4,...,4}{
        \gate{(-\doubledx*\x-\doubledx*0.5,\doubledx/2*\x)}{\scriptsize $U_{N}$};          
        }
        \end{scope}
        \end{tikzpicture}
                \end{array}
\end{align}
with all $U_k$ unitary, which is a MPU by identifying
\begin{align} \label{eq:1_floor_tensors}
        \begin{array}{c}
        \begin{tikzpicture}[scale=0.6,baseline={([yshift=-0.65ex] current bounding box.center)}]
            \ATensor{0,0}{\small $A_N$}{-1}
        \end{tikzpicture}
        =
         \begin{tikzpicture}[scale=0.6,baseline={([yshift=-.7ex] current bounding box.center)}]
                \draw [thick] (0.7,0.7) -- (0.7,-.7) -- (1.2,-.7);
                \draw [thick] (-0.7,-0.7) -- (-0.7,.7); 
                \gate{0,0}{\scriptsize $U_{N}$};
            \end{tikzpicture}
, \quad  
        \begin{tikzpicture}[scale=0.6,baseline={([yshift=-0.65ex] current bounding box.center)}]
            \ATensor{0,0}{\small $A_k$}{0}
        \end{tikzpicture}
        =
         \begin{tikzpicture}[scale=0.6,baseline={([yshift=-.7ex] current bounding box.center)}]
                \draw [thick] (-0.7,-0.7) -- (-0.7,.7) -- (-1.2,.7);
                \draw [thick] (0.7,0.7) -- (0.7,-.7) -- (1.2,-.7);
                \gate{0,0}{\scriptsize $U_{k}$};
            \end{tikzpicture}
, \quad
        \begin{tikzpicture}[scale=0.6,baseline={([yshift=-0.65ex] current bounding box.center)}]
            \ATensor{0,0}{\small $A_1$}{1}
        \end{tikzpicture}
        =
         \begin{tikzpicture}[scale=0.6,baseline={([yshift=-.7ex] current bounding box.center)}]
                \draw [thick] (-0.7,-0.7) -- (-0.7,.7) -- (-1.2,.7);
                \draw [thick] (0.7,-0.7) -- (0.7,.7); 
                \gate{0,0}{\scriptsize $U_{1}$};
            \end{tikzpicture}
        \end{array}
        .
\end{align}
Unitarity of the gates implies that the MPU is automatically in canonical form. Moreover $\rho_k = \1/D_k$ and $\mathcal S^{(k)}$ contains all traceless Hermitian operators. This last fact follows because for all $X \in \mathcal S^{(k)}$,
\begin{align}
            \begin{array}{c}
            \begin{tikzpicture}[scale=0.6,baseline={([yshift=-0.65ex] current bounding box.center)}]
        		      \foreach \x in {0,...,0}{
            \SideIdentityTensor{\x*\doubledx,0}{\small $X$}{\stradius}{3};
            }
            \end{tikzpicture}
            \end{array}
     = 
        \begin{array}{c}
    \begin{tikzpicture}[scale=0.6]
        \draw[thick, fill=whitetensorcolor, rounded corners=2pt] (-3.7*\doubledx,2.8) rectangle (0.5*\doubledx,1.8);
	    \draw (-2*\doubledx,2.3) node {\small $\widetilde X$};
		      \foreach \x in {0,...,0}{
                \DoubleLongLine{-\doubledx*\x,0}
                \DoubleATensor{(-\doubledx*\x,0)}{1}
                \draw (-\doubledx*\x,-0.8) node {\small $A_{1}$};
        }
		      \foreach \x in {1,...,1}{
                \DoubleLongLine{-\doubledx*\x,0}
                \DoubleATensor{(-\doubledx*\x,0)}{0}
                \DoubleATensor{(-\doubledx*\x,0)}{1}
                \draw (-\doubledx*\x,-0.8) node {\small $A_{2}$};
        }
		      \foreach \x in {2,...,2}{
                \DoubleDots{-\doubledx*\x,0}{\doubledx/2}
        }
		      \foreach \x in {3,...,3}{
                \DoubleLongLine{-\doubledx*\x,0}
                \DoubleATensor{(-\doubledx*\x,0)}{0}
                \DoubleATensor{(-\doubledx*\x,0)}{1}
                \draw (-\doubledx*\x,-0.8) node {\small $A_{k}$};
        }
    \end{tikzpicture}
    \end{array}
\end{align}
where $\widetilde X = U^\dagger_1 \dots U^\dagger_k X_k U_k \dots U_1$ is also Hermitian and traceless.

The converse also holds true. That is, every MPU in canonical form such that $\rho_k = \1/D_k$ and $\mathcal S^{(k)}$ contains all traceless Hermitian operators is a 1-floor staircase circuit. That is, it has the form~\eqref{eq:1_floor_staicase} using the identification~\eqref{eq:1_floor_tensors}. To show this, notice that \cref{eq:inhom_unitarity_local} implies \cref{eq:local_isometry}. This itself indicates that all tensors $A_k$ are isometries, with output space (top and left legs) at least at large as input one (bottom and right legs). However, even if a single tensor is not unitary, then necessarily the global operator $U$ is not a unitary but an isometry, which violates the initial assumption.

Notice that $A_1$ and $A_N$ have unequal input-output physical dimensions. This circuit construction thus shows how the no-go result in \cref{ex:depolarizing} can be circumvented by allowing unequal input-output dimensions of the tensors.
\end{example}


It is easy to see that the sequential circuit cannot express every MPU with $D=d$. The reason can be traced back to causality due to the orientation of the staircase (right to left in \cref{ex:1_floor}). For instance, given any local observable $O_i$, the time-evolved $U^\dagger O_i U$ is not supported over sites to the left of $i$, except its immediate neighbor. Clearly, this property fails in general for a staircase with the opposite orientation, which is also a valid MPU of the same bond dimension.

This motivates searching for more general architectures that are in canonical form:

\begin{example}[2-floor staircase] \label{ex:2_floor}
Consider the family of quantum circuits
\begin{align} \label{eq:2_floor_staicase}
                 \begin{array}{c}
        \begin{tikzpicture}[scale=.6]
		      \foreach \x in {1,...,5}{
      \draw [thick] (-\doubledx*\x,-3.4) -- (-\doubledx*\x,5);
        }
		\foreach \x in {1,...,2}{
        \gate{(-\doubledx*\x-\doubledx*0.5,\doubledx/2*\x)}{\scriptsize $U_\x$};          
        }
		\fill[fill=white] (-3.1*\doubledx,2.8) rectangle (-2.9*\doubledx,2.2);
        \draw [very thick, dotted] (-2.8*\doubledx,2.2) to (-3.2*\doubledx,2.8);
        \begin{scope}[shift={(0,1)}]
		\foreach \x in {3,...,3}{
        \gate{(-\doubledx*\x-\doubledx*0.5,\doubledx/2*\x)}{\scriptsize $U_{N-1}$};          
        }
		\foreach \x in {4,...,4}{
        \gate{(-\doubledx*\x-\doubledx*0.5,\doubledx/2*\x)}{\scriptsize $U_{N}$};          
        }
        \end{scope}
    \begin{scope}[shift={(0,1.6)}]
		\foreach \x in {2,...,2}{
        \gate{(-\doubledx*\x-\doubledx*0.5,-\doubledx/2*\x)}{\scriptsize $V_\x$};          
        }
		\fill[fill=white] (-3.1*\doubledx,-2.8) rectangle (-2.9*\doubledx,-2.2);
        \draw [very thick, dotted] (-2.8*\doubledx,-2.2) to (-3.2*\doubledx,-2.8);
        \begin{scope}[shift={(0,-1)}]
		\foreach \x in {3,...,3}{
        \gate{(-\doubledx*\x-\doubledx*0.5,-\doubledx/2*\x)}{\scriptsize $V_{N-1}$};          
        }
		\foreach \x in {4,...,4}{
        \gate{(-\doubledx*\x-\doubledx*0.5,-\doubledx/2*\x)}{\scriptsize $V_{N}$};          
        }
        \end{scope}
    \end{scope}
        \end{tikzpicture}
                \end{array}
\end{align}
with all $U_k, V_k$ unitary, which is a MPU by identifying
\begin{align} \label{eq:2_floor_tensors}
        \begin{array}{c}
        \begin{tikzpicture}[scale=0.6,baseline={([yshift=-0.65ex] current bounding box.center)}]
            \ATensor{0,0}{\small $A_N$}{-1}
        \end{tikzpicture}
        =
         \begin{tikzpicture}[scale=0.6,baseline={([yshift=-.7ex] current bounding box.center)}]
            \begin{scope}[shift={(0,.9)}]
                \draw [thick] (-0.7,-0.7) -- (-0.7,.7);
                \draw [thick] (0.7,0.7) -- (0.7,-.7) -- (1.2,-.7);
                \gate{0,0}{\scriptsize $U_{N}$};
            \end{scope}
            \begin{scope}[shift={(0,-.9)}]
                \draw [thick] (-0.7,1.2) -- (-0.7,-.7);
                \draw [thick] (1.2,.7) -- (0.7,0.7) -- (0.7,-.7);
                \gate{0,0}{\scriptsize $V_{N}$};
            \end{scope}
            \end{tikzpicture}
\;, \quad 
        \begin{tikzpicture}[scale=0.6,baseline={([yshift=-0.65ex] current bounding box.center)}]
            \ATensor{0,0}{\small $A_k$}{0}
        \end{tikzpicture}
        =
         \begin{tikzpicture}[scale=0.6,baseline={([yshift=-.7ex] current bounding box.center)}]
            \begin{scope}[shift={(0,.9)}]
                \draw [thick] (-0.7,-0.7) -- (-0.7,.7) -- (-1.2,.7);
                \draw [thick] (0.7,0.7) -- (0.7,-.7) -- (1.2,-.7);
                \gate{0,0}{\scriptsize $U_{k}$};
            \end{scope}
            \begin{scope}[shift={(0,-.9)}]
                \draw [thick] (-0.7,1.2) -- (-0.7,-.7) -- (-1.2,-.7);
                \draw [thick] (1.2,.7) -- (0.7,0.7) -- (0.7,-.7);
                \gate{0,0}{\scriptsize $V_{k}$};
            \end{scope}
            \end{tikzpicture}
\;, \quad
        \begin{tikzpicture}[scale=0.6,baseline={([yshift=-0.65ex] current bounding box.center)}]
            \ATensor{0,0}{\small $A_1$}{1}
        \end{tikzpicture}
        =
         \begin{tikzpicture}[scale=0.6,baseline={([yshift=-.7ex] current bounding box.center)}]
                \draw [thick] (-1.2,-0.7) -- (-0.7,-0.7) -- (-0.7,.7) -- (-1.2,.7);
                \draw [thick] (0.7,-0.7) -- (0.7,.7); 
                \gate{0,0}{\scriptsize $U_{1}$};
            \end{tikzpicture}
        \end{array}
        .
\end{align}
It is easy to verify directly the representation \cref{eq:2_floor_tensors} is in canonical form with $D = d^2$.
\end{example}

The architecture of \cref{ex:2_floor} is more general and violates the causality constraint of its 1-floor counterpart. Nevertheless, we now show that this family cannot express every MPU with $D = d^2$.

\begin{restatable}[]{proposition}{doublestaircase} \label{prop:doublestaircase}
    Given $N \ge 2$ and an arbitrary set of 2-qubit unitaries $\widetilde {U}_k,\widetilde {V}_k$ it is \textbf{not} always possible to find another set of 2-qubit unitaries $U_k,V_k$ such that 
    \begin{align} \label{eq:app:left-right-staircase}
        \begin{array}{c}
        \begin{tikzpicture}[scale=.7]
		      \foreach \x in {1,...,4}{
      \draw [thick] (\doubledx*\x,-2.4) -- (\doubledx*\x,4);
        }
		\foreach \x in {1,...,2}{
        \gate{(\doubledx*\x+\doubledx*0.5,\doubledx/2*\x)}{\tiny $\widetilde{U}_\x$};          
        }
		\fill[fill=white] (3.1*\doubledx,2.8) rectangle (2.9*\doubledx,2.2);
        \draw [very thick, dotted] (2.8*\doubledx,2.2) to (3.2*\doubledx,2.8);
        \begin{scope}[shift={(0,1)}]
		\foreach \x in {3,...,3}{
        \gate{(\doubledx*\x+\doubledx*0.5,\doubledx/2*\x)}{\tiny $\widetilde{U}_{N}$};          
        }
        \end{scope}
    \begin{scope}[shift={(0,1.6)}]
		\foreach \x in {2,...,2}{
        \gate{(\doubledx*\x+\doubledx*0.5,-\doubledx/2*\x)}{\tiny $\widetilde{V}_\x$}; 
        }
		\fill[fill=white] (3.1*\doubledx,-2.8) rectangle (2.9*\doubledx,-2.2);
        \draw [very thick, dotted] (2.8*\doubledx,-2.2) to (3.2*\doubledx,-2.8);
        \begin{scope}[shift={(0,-1)}]
		\foreach \x in {3,...,3}{
        \gate{(\doubledx*\x+\doubledx*0.5,-\doubledx/2*\x)}{\tiny $\widetilde{V}_{N}$};          
        }
        \end{scope}
    \end{scope}
        \end{tikzpicture}
                \end{array}
            =
                 \begin{array}{c}
        \begin{tikzpicture}[scale=.7]
		      \foreach \x in {1,...,4}{
      \draw [thick] (-\doubledx*\x,-2.4) -- (-\doubledx*\x,4);
        }
		\foreach \x in {1,...,2}{
        \gate{(-\doubledx*\x-\doubledx*0.5,\doubledx/2*\x)}{\tiny $U_\x$};          
        }
		\fill[fill=white] (-3.1*\doubledx,2.8) rectangle (-2.9*\doubledx,2.2);
        \draw [very thick, dotted] (-2.8*\doubledx,2.2) to (-3.2*\doubledx,2.8);
        \begin{scope}[shift={(0,1)}]
		\foreach \x in {3,...,3}{
        \gate{(-\doubledx*\x-\doubledx*0.5,\doubledx/2*\x)}{\tiny $U_{N}$};          
        }
        \end{scope}
    \begin{scope}[shift={(0,1.6)}]
		\foreach \x in {2,...,2}{
        \gate{(-\doubledx*\x-\doubledx*0.5,-\doubledx/2*\x)}{\tiny $V_\x$};          
        }
		\fill[fill=white] (-3.1*\doubledx,-2.8) rectangle (-2.9*\doubledx,-2.2);
        \draw [very thick, dotted] (-2.8*\doubledx,-2.2) to (-3.2*\doubledx,-2.8);
        \begin{scope}[shift={(0,-1)}]
		\foreach \x in {3,...,3}{
        \gate{(-\doubledx*\x-\doubledx*0.5,-\doubledx/2*\x)}{\tiny $V_{N}$};          
        }
        \end{scope}
    \end{scope}
        \end{tikzpicture}
                \end{array}
                .
\end{align}
\end{restatable}
The proof can be found in \cref{sec:app:proofs_inhomo_canonical}.

Finally, we note that a decomposition similar to \cref{ex:2_floor} in the context of entanglement theory is also considered in Ref.~\cite{pozsgay2024tensor}.

\section{MPU with uniform bulk and arbitrary boundary} \label{sec:homo}

In \cref{sec:motivating_example} we identified a TI MPU with open boundary, namely the multi-control $Z$-gate, which is not a QCA. Here we explore this class more systematically. In particular, we are interested in MPO specified by a single bulk tensor $A$ (independent of $N$) and a boundary condition $b$ such that
\begin{align} \label{eq:homo_unitarity_def}
    U_N = \sum_{i,j} \Tr \left( b A^{i_1 j_1} \dots  A^{i_N j_N} \right) \ket{i_1 \dots i_N}  \bra{j_1 \dots j_N}
\end{align}
is unitary for all $N > 1$.

Depending on the choice of the boundary $b$, the resulting operator $U_N$ may or may not be TI. In general, a uniform MPO specified by a tensor $A$ is TI for all $N$ if and only if
\begin{align}
    \Tr \left( b [X, Y] \right) = 0 \quad \forall X,Y \in \mathrm{Alg} (A)
\end{align}
where $\mathrm{Alg} (A)$ is the algebra generated by $\{ A^{ij} \}_{i,j=1}^d$~\cite{marta2024private}.

Our starting point is to formulate unitarity over the auxiliary space. This directly follows as a special case of \cref{prop:unitarity_bond_S}. Since the bulk of the tensor network is uniform, we henceforth drop the site labels.
\begin{lemma}  \label{lem:unitarity_bond_S_homo}
    The tensor $A$ with boundary conditions $b$ generates a MPU over $N$ sites [cf.~\cref{eq:homo_unitarity_def}] if and only if
    \begin{subequations} \label{eq:homo_unitarity_bond}
    \begin{numcases}{\Tr \left(B S_{j_N} \dots S_{j_2} S_{j_1} \right) = }
           \label{eq:homo_unitarity_bond_1}  1 & if  all $j_k=0$, \\
            \label{eq:homo_unitarity_bond_2} 0 & otherwise.
    \end{numcases}
    \end{subequations}
where $B = b^* \otimes b$.
\end{lemma}

We will soon need the notion of blocking~\cite{cirac2021matrix}. This just amounts to combining $q$ tensors together
\begin{align}
    \begin{array}{c}
        \begin{tikzpicture}[scale=0.5,baseline={([yshift=-0.8ex] current bounding box.center)}]
            \draw[thick,dotted] (0,0) to (2,0);
            \ATensor{(0,0)}{}{0};
            \ATensor{(3,0)}{}{0};
        \end{tikzpicture}
    \end{array}
    \mapsto
            \begin{array}{c}
            \begin{tikzpicture}[scale=0.5,baseline={([yshift=-1ex] current bounding box.center)}]
                \begin{scope}
		              \draw[very thick] (-1.4,0) -- (1.4,0);
		              \draw[very thick] (-0.6,-1) -- (-0.6,1);
                    \draw[very thick] (0.6,-1) -- (0.6,1);
                    \draw[dotted, thick] (-0.3,.85) -- (0.3,.85);
                    \draw[dotted, thick] (-0.3,-.85) -- (0.3,-.85);
                    \draw[thick, fill=tensorcolor, rounded corners=2pt] (-1,-0.6) rectangle (1,0.6);
	            \end{scope}
            \end{tikzpicture}
            \end{array}
\end{align}
to a new tensor with physical dimension $d^q$ and the same bond dimension. As it is often done, we allow blocking a finite number of times (i.e., independent of $N$) as this freedom eliminates irrelevant short-ranged features.

One of the difficulties in dealing with open-boundary conditions is that, in contrast to periodic, it might not always be possible to find a representation such that $A^{ij}$ can be taken to be a direct sum of ``simple" objects. For instance, the canonical form of MPS with periodic boundary asserts that, after blocking, any tensor can be written as a direct sum of normal\footnote{A tensor is called injective if the corresponding linear map from auxiliary to physical space is injective. A tensor is normal if it becomes injective after blocking~\cite{cirac2021matrix}.} tensors~\cite{cirac2017matrix1}. That is, all uniform MPS can be chosen (in canonical form) to have a block diagonal structure. However, this fails in general when considering MPS with open boundary\footnote{An example is the $W$ state in its usual $D=2$ MPS representation~\cite{cirac2021matrix}.}.

For MPU, we will see that a similar conclusion holds. That is, in general, MPU with uniform bulk but open boundary do not admit a block-diagonal decomposition. Nevertheless, there are interesting examples with block-diagonal structure, such as the multi-control $Z$-gate of \cref{ex:multi_control_z}, and its generalization:

\begin{example}[Unitary action over product subspace] \label{ex:subspace_product}
    Given any $V \in U(2)$, define the $N$-qubit unitary
    \begin{align}
        U = \1^{\otimes N} + \sum_{i,j=0}^1 (V_{ij} - \delta_{ij}) \ket{i}^{\otimes N} \bra{j}^{\otimes N}
    \end{align}
    which acts as $V$ over the subspace $\Span \{\ket{0}^{\otimes N},\ket{1}^{\otimes N}\}$ and trivially otherwise. $U$ can be written as a uniform $D = 5$ MPU
    \begin{subequations}
    \begin{align}
        A^{00} = \diag ( 1,1,0,0,0) \\
        A^{11} = \diag ( 1,0,1,0,0) \\
        A^{01} = \diag ( 0,0,0,1,0) \\
        A^{10} = \diag ( 0,0,0,0,1)
    \end{align}
    \end{subequations}
where the action of $V$ is encoded in the boundary condition
\begin{align}
    b = \diag(1,V_{00} - 1, V_{11} - 1, V_{01}, V_{10} )  \,.
\end{align}
The resulting MPU is clearly block-diagonal and also permutation invariant, because ${\mathrm {Alg}}(\{ A^{ij}\}_{ij}\cup\{ b\} )$ is abelian.

The construction includes as a special case the multi-control $Z$-gate (\cref{ex:multi_control_z}). There a part of the boundary vanishes thus ones can compress to $D=2$. It also generalizes directly to arbitrary physical dimension $d$, with $D = d^2+1$.
\end{example}

In the remaining of this section we focus on the subclass that admits a direct-sum decomposition, which we call semi-simple MPU.

\subsection{Semi-simple MPU}

The main idea is to assume a direct-sum form not over the tensor $A^{ij}$, but instead over the span
\begin{align}
    \mathcal A = \Span \{ S_{\alpha} \}_{\alpha= 0}^{d^2-1}
\end{align}
(recall that $E \equiv S_0$ corresponds to the transfer matrix).

\begin{definition}
    A MPU is semi-simple if it admits a representation with $\mathcal A = \bigoplus_i \mathcal A_i$ where each $\mathcal A_i$ is irreducible.
\end{definition}
Note that the direct sum structure may be obtained via similarity transformations acting jointly over $\mathbb C^D \otimes \mathbb C^D$. We remind the reader that a set of operators is irreducible if its elements do not have any (nontrivial) common invariant subspace. We will now show that the semi-simplicity assumption strongly restricts the form of the resulting MPU. For this, we need to first recall some useful facts for irreducible subspaces.

The irreducibility of a block $\mathcal A_i$ does not necessarily imply that $\mathcal A_i$ is the full matrix algebra. For example, $\mathcal A = \Span \{\sigma^+,\sigma^-\}$ is irreducible but not closed under multiplication ($\sigma^\pm \propto X \pm i Y$). Nevertheless, notice that blocking twice results in $\mathcal A^2 = \Span\{\ket{0}\bra{0}, \ket{1}\bra{1}\}$ which is a direct sum of (1-dimensional) full matrix algebras.

In fact, every span $\mathcal A = \bigoplus_i \mathcal A_i$ with $\mathcal A_i$ irreducible can be expressed, after blocking sufficiently many times, as a direct sum of full matrix algebras, as in the example above~\cite{cirac2017matrix1}. The resulting algebras may be taken to be either proportional to each other or independent~\cite{perez_garcia2007matrix}. That is, the span $\mathcal A$ of a semi-simple MPU can be brought by a similarity transformation to the so-called block-injective canonical form~\cite{cirac2017matrix1}
\begin{align} \label{eq:homo_block_injective}
 \mathcal A = \bigoplus_j \diag (\lambda_{j1},\dots, \lambda_{jm_j}) \otimes \mathbb M_{d_j}
\end{align}
where $\mathbb M_{d_j}$ denotes the full matrix algebra of dimension $d_j$ and $\lambda_{jk} \in \mathbb C$.

In words, this means that after blocking and performing a similarity transformation, a basis for the span $\mathcal A$ can be taken to consist of a direct sum of square matrices. Crucially, the different matrices comprising an element in the basis are either proportional (matrices within a fixed $j$) or can be chosen independently (matrices corresponding to distinct values of $j$).

\begin{example}[Multi-control $Z$-gate is semi-simple] Using the representation \cref{eq:tensors_cz}, the multi-control $Z$-gate [cf. \cref{eq:controlZ}] corresponds to
\begin{subequations}
\begin{align}
    E &= \diag (1,1/2,1/2,1/2) \\
    S_3 &= \diag (0,-1/2,-1/2,-1/2) \\
    S_1 &= S_2 = 0 \\
    B &= \diag(1,-2,-2,4) + \text{off-diagonal}
\end{align}
\end{subequations}
(matrix representation in the product basis and $S_i$ corresponds to Pauli $\sigma_i$). Thus $\mathcal A$ takes the form \eqref{eq:homo_block_injective} with $d_1 = d_2 = 1$ and multiplicities $m_1=1$, $m_2 = 3$. 
\end{example}

\begin{example}[Control-$X$ staircase is not semi-simple] Consider the 1-floor staircase of \cref{eq:1_floor_staicase} with $d=2$ where all gates are control-$X$ with control on the right. This can be expressed as a uniform $D=2$ MPU with
\begin{subequations}
\begin{align}
        \begin{array}{c}
        \begin{tikzpicture}[scale=0.6,baseline={([yshift=-0.65ex] current bounding box.center)}]
            \ATensor{0,0}{\small $A$}{0}
        \end{tikzpicture}
        =
         \begin{tikzpicture}[scale=0.6,baseline={([yshift=-.7ex] current bounding box.center)}]
                \draw [thick] (0,-1) -- (0,1);
                \draw [thick] (0,-0.3) -- (.8,-.3);
                \draw [thick] (0,.6) -- (-.6,.6);
		\fill[color=black] (0,0.6) circle (0.12);
         \begin{scope}[shift={(0,-0.3)}]
             \draw[ thick, fill=whitetensorcolor, rounded corners=2pt] (-.4,-.4) rectangle (0.4,0.4); 
	        \draw (0,0) node {\small $X$};
         \end{scope}
            \end{tikzpicture}
        \end{array}
\end{align}
and boundary conditions
    \begin{align}
            \bra{l} = \bra{0}+ \bra{1} \;, \quad \ket{r} = \ket{0}  \,.
        \end{align}
    \end{subequations}
Then, in the (normalized) Pauli basis,
    \begin{align}
        E = \ket{\1}\bra{\1}, \quad 
        S_X = \ket{\1}\bra{X}, \quad
        S_Y = \ket{Z}\bra{Y}, \quad
        S_Z = \ket{Z}\bra{Z} \,.
    \end{align}
Thus, in this tensor-network representation, the MPU is not semi-simple. This is because blocking leaves $\mathcal A$ unchanged while the off-diagonal contribution of $S_X, S_Y$ cannot be removed by a similarity transformation.
\end{example}

%
%

\subsection{Structure of semi-simple MPU}

We now refine the form of \cref{eq:homo_block_injective} by imposing unitarity. As in the nonuniform case, we make the distinction between the span of the traceless operators
\begin{align}
    \mathcal S = \Span\{S_\alpha\}_{\alpha=1}^{d^2-1}
\end{align}
and $\mathcal A = \Span( {\mathcal S}\cup \{E\})$ where $E \equiv S_0$ corresponds to the usual transfer matrix.
\begin{proposition}[Structure of semi-simple MPU]\label{prop:structure_semisimple}
    After blocking a finite number of times, every semi-simple MPU satisfies
    \begin{align}
        \mathcal S E = E \mathcal S \subseteq \mathcal S ^2 \,.
    \end{align}
    Up to a similarity transformation over $\mathbb C^D \otimes \mathbb C^D$,
    \begin{subequations}
        \begin{alignat}{3}
            E &= \1  && \oplus Q_E && {}\\
            \mathcal S &= 0 &&\oplus Q_S && {} \\
            B &= B_{\1} &&\oplus B_Q && + \text{off-diagonal.}
        \end{alignat}
    \end{subequations}
    where
\begin{subequations}
\begin{align}
            Q_E &= \bigoplus_j \diag(\lambda_{j1},\dots, \lambda_{jm_j} ) \otimes E_j   \\
            Q_S &= \bigoplus_j \diag(\lambda_{j1},\dots, \lambda_{jm_j} ) \otimes \mathbb M_{d_j}
\end{align}
\end{subequations}
with $\Tr(B_\1) = 1$ and $\Tr(B_Q Q_E) = \Tr(B_Q Q_S)= 0$.
\end{proposition}

The proof can be found in \cref{sec:app:proofs_homo}.

Let us unpack the above structure. In short, it asserts that every semi-simple MPU splits in a direct sum of two blocks. Over the first block, the transfer matrix acts as identity, which leads to the unit contribution in \cref{eq:homo_unitarity_bond_1}, while $\mathcal S$ vanishes. The second block, although possibly non-vanishing, does not contribute to the trace in \cref{eq:homo_unitarity_bond}.

A consequence of the above structure is that semi-simple MPU with periodic boundary are necessarily product unitaries.

\begin{corollary}
    After blocking a finite number of times, semi-simple MPU with periodic boundary conditions factorize, i.e.,
    \begin{align} \label{eq:semisimple_factorization}
            U = e^{i \theta N} V^{\otimes N} \quad \forall N
    \end{align}
    where $V$ is unitary.
\end{corollary}
\begin{proof}
    Since $B_Q = \1$ we have $\Tr(Q^N_S) = 0$ $\forall N$ (because we have an MPU for all $N$). This is only possible for $\lambda_{jk} = 0$ (see \cref{app:lemma:vandermonde} for a proof) and  thus $\mathcal S = 0$, which implies \cref{eq:semisimple_factorization}.
\end{proof}
\noindent Thus the semi-simplicity assumption is never satisfied for QCA when the latter are represented as homogeneous MPU with periodic boundary~\cite{cirac2017matrix2} (except from the trivial case of product unitaries). This can also be deduced directly from the proof in Ref.~\cite{cirac2017matrix1} since, after blocking, QCA satisfy $\mathcal S ^3 = 0$.

%


We finish the section with an example of a TI family, which is not necessarily permutation invariant. The idea is to have $V$ of \cref{ex:subspace_product} act over a subspace which, instead of only product states, is spanned by MPS renormalization-group (RG) fixed-points. Recall that, for normal tensors, the latter can be represented (for uniform MPS) as
\begin{align} \label{eq:rg_fp}
        \begin{array}{c}
            \begin{tikzpicture}[scale=0.6,baseline={([yshift=1.5ex] current bounding box.center)}]
		\draw[very thick] (0,0) -- (0,1);
		\draw[very thick] (-1,0) -- (1,0);
        \draw[thick, fill=tensorcolor, rounded corners=2pt] (-0.5,-0.5) rectangle (0.5,0.5);
		\draw (0,0) node {\small $B_i$};
            \end{tikzpicture}
        \end{array}
    =
        \begin{array}{c}
         \begin{tikzpicture}[scale=0.6,baseline={([yshift=-.2ex] current bounding box.center)}]
                \draw [very thick] (-.8,-.2) -- (0,-.2) -- (0,1.6);
                \draw [very thick] (.8,1.6) -- (.8,-.2) -- (1.6,-.2);\filldraw[thick,color=black,fill=whitetensorcolor] (0,0.7) circle (.58);
	            \draw (-.05,0.7) node {\small $\sqrt{\rho_i}$};
            \end{tikzpicture}
        \end{array}
\end{align}
where $\rho_i$ can be any full-rank density matrix~\cite{verstraete2005renormalization,cirac2017matrix2}.
Thus two RG fixed-points are orthogonal (as states) if $\Tr(\sqrt{\rho_i} \sqrt{\rho_j}) = 0$.

\begin{example}[Unitary action over RG subspaces] \label{ex:rg_subspace}
Consider a set of mutually orthogonal RG fixed-point $\{ \ket{\Psi_i} \bra{\Psi_i} \}_{i=1}^r$ of bond dimension $D_i$ and a unitary $V \in U(r)$. Then
    \begin{align}
        U = \1^{\otimes N} + \sum_{i,j=1}^r (V_{ij} - \delta_{ij}) \ket{\Psi_i} \bra{\Psi_j}
    \end{align}
is unitary. Moreover, $U - \1^{\otimes N}$ can be expressed as a MPO with
\begin{align}
        \begin{array}{c}
            \begin{tikzpicture}[scale=0.6,baseline={([yshift=0.5ex] current bounding box.center)}]
		\draw[very thick] (0,-.8) -- (0,.8);
		\draw[very thick] (-.8,0) -- (.8,0);
        \draw[thick, fill=tensorcolor, rounded corners=2pt] (-0.5,-0.5) rectangle (0.5,0.5);
		\draw (0,0) node {\small $A$};
            \end{tikzpicture}
        \end{array}
    =
        \begin{array}{c}
         \begin{tikzpicture}[scale=0.6,baseline={([yshift=1.5ex] current bounding box.center)}]
        \begin{scope}[shift={(0,-.5)}]
                \draw [very thick] (-1,0.8) -- (1,0.8);
                \draw [very thick] (-1,1.6) -- (1,1.6);
                \draw [very thick] (0,2.5) -- (0,0.8);
        \draw[thick, fill=tensorcolor, rounded corners=2pt] (.5,2.1) rectangle (-0.5,1.1);
		\draw (0,1.6) node {\small $B$};
		\fill[color=black] (0,0.8) circle (0.12);
        \end{scope}
        \begin{scope}[shift={(0,+.5)}]
                \draw [very thick] (-1,-0.8) -- (1,-0.8);
                \draw [very thick] (-1,-1.6) -- (1,-1.6);
                \draw [very thick] (0,-2.5) -- (0,-0.8);
        \draw[thick, fill=tensorcolor, rounded corners=2pt] (.5,-2.1) rectangle (-0.5,-1.1);
		\draw (0,-1.6) node {\small ${B^*}$};
		\fill[color=black] (0,-0.8) circle (0.12);
        \end{scope}
            \end{tikzpicture}
        \end{array}
        \,, \quad   
    b =
        \begin{array}{c}
            \begin{tikzpicture}[scale=0.6,baseline={([yshift=0.5ex] current bounding box.center)}]
		\draw[very thick] (-1,0.8) -- (1,.8);
		\draw[very thick] (-1,-0.8) -- (1,-.8);
		\draw[very thick] (-1,1.2) -- (1,1.2);
		\draw[very thick] (-1,-1.2) -- (1,-1.2);
		\draw[very thick] (0,-0.8) -- (0,0.8);
    	\filldraw[color=black, fill=whitetensorcolor, thick] (0,0) ellipse (1 and 0.5);
	\draw (0,0) node {\small $V - \1$};
		\fill[color=black] (0,0.8) circle (0.12);
		\fill[color=black] (0,-0.8) circle (0.12);
            \end{tikzpicture}
        \end{array}
    \,, \quad
        \begin{array}{c}
            \begin{tikzpicture}[scale=0.6,baseline={([yshift=5ex] current bounding box.center)}]
		\draw[very thick] (0,-.8) -- (0,.8);
		\draw[very thick] (-.8,0) -- (.8,0);
        \draw[thick, fill=tensorcolor, rounded corners=2pt] (-0.5,-0.5) rectangle (0.5,0.5);
		\draw (0,0) node {\small $B$};
		\draw (0,-1.2) node {\small $i$};
            \end{tikzpicture}
        \end{array}
    \equiv
        \begin{array}{c}
            \begin{tikzpicture}[scale=0.6,baseline={([yshift=-1.5ex] current bounding box.center)}]
		\draw[very thick] (0,0) -- (0,.8);
		\draw[very thick] (-.8,0) -- (.8,0);
        \draw[thick, fill=tensorcolor, rounded corners=2pt] (-0.5,-0.5) rectangle (0.5,0.5);
		\draw (0,0) node {\small $B_i$};
            \end{tikzpicture}
        \end{array}
\end{align}
where each tensor $B_i$ corresponds to the fixed-point MPS $\ket{\Psi_i}$. Thus $U$ can be represented with $D = ( \sum D_i)^2 + 1$. Note that this family includes both normal and non-normal tensors, as well as the possibility for a non-abelian span $\mathcal A$.

Moreover $U$ is a semi-simple MPU in the above representation. This follows by a direct calculation of $\mathcal A$ by using that (i) the different $B_{i}$ tensors are locally orthogonal after blocking twice [see \cref{eq:rg_fp}], and (ii) $A$ is irreducible since $B$ is in block-injective canonical form~\cite{cirac2017matrix2}.
\end{example}

\section{MPU and locally maximally-entanglable states} \label{sec:lme}

In this section we observe that vectorizing a multipartite unitary and then locally isometrically compressing the state results in a so-called locally maximally entanglable (LME) state. The latter is a class previously introduced in multipartite entanglement theory~\cite{kruszynska2009local}. Via this correspondence we show that all multi-qubit unitaries whose LME state can be compressed to local dimension $d' = 2$ are local-unitary (LU) equivalent to unitaries diagonal in a product basis. This class includes control gates with multiple control qubits.

\subsection{Multipartite unitaries and LME states}


Recall that vectorizing a unitary
\begin{align}
    \ket{U}_{AB} = \sum_{ij} U_{ij} \ket{i}_A \ket{j}_B
\end{align}
results in an (unnormalized) maximally entangled state, also known as the Choi-Jamio{\l}kowski state~\cite{nielsen2002quantum}. In the spirit of tensor networks~\cite{verstraete2005renormalization,cirac2017matrix2}, we now consider the locally compressed $\ket{U}_{AB}$. That is, we express
    \begin{align} \label{eq:compressed_CJ}
        \ket{U}_{AB} = \bigotimes_k V_k \ket{\phi_U} =
    \begin{array}{c}
    \begin{tikzpicture}[scale=.6]
		      \foreach \x in {0,...,0}{
        \draw[very thick] (-\x*\doubledx,0) -- (-\x*\doubledx,1.5);
        \draw[very thick] (-\x*\doubledx+0.3,2.5) -- (-\x*\doubledx+0.3,3);
        \draw[very thick] (-\x*\doubledx-0.3,2.5) -- (-\x*\doubledx-0.3,3);
        \draw[thick, fill=whitetensorcolor, rounded corners=2pt] (-\x*\doubledx+.5,1.5+1) rectangle (-\x*\doubledx-.5,1.5); 
        \draw (-\x*\doubledx,2) node {\small $V_1$};
        \draw (-\x*\doubledx+0.4,3.3) node {\scriptsize $A_1$};
        \draw (-\x*\doubledx-0.4,3.3) node {\scriptsize $B_1$};
        }
		      \foreach \x in {1,...,1}{
        \draw[very thick] (-\x*\doubledx,0) -- (-\x*\doubledx,1.5);
        \draw[very thick] (-\x*\doubledx+0.3,2.5) -- (-\x*\doubledx+0.3,3);
        \draw[very thick] (-\x*\doubledx-0.3,2.5) -- (-\x*\doubledx-0.3,3);
        \draw[thick, fill=whitetensorcolor, rounded corners=2pt] (-\x*\doubledx+.5,1.5+1) rectangle (-\x*\doubledx-.5,1.5); 
        \draw (-\x*\doubledx,2) node {\small $V_2$};
        \draw (-\x*\doubledx+0.4,3.3) node {\scriptsize $A_2$};
        \draw (-\x*\doubledx-0.4,3.3) node {\scriptsize $B_2$};
        }
		      \foreach \x in {2,...,2}{
        \draw[very thick, dotted] (-\x*\doubledx-0.5,2) -- (-\x*\doubledx+0.5,2);
        }
		      \foreach \x in {3,...,3}{
        \draw[very thick] (-\x*\doubledx,0) -- (-\x*\doubledx,1.5);
        \draw[very thick] (-\x*\doubledx+0.3,2.5) -- (-\x*\doubledx+0.3,3);
        \draw[very thick] (-\x*\doubledx-0.3,2.5) -- (-\x*\doubledx-0.3,3);
        \draw[thick, fill=whitetensorcolor, rounded corners=2pt] (-\x*\doubledx+.5,1.5+1) rectangle (-\x*\doubledx-.5,1.5); 
        \draw (-\x*\doubledx,2) node {\small $V_N$};
        \draw (-\x*\doubledx+0.4,3.3) node {\scriptsize $A_N$};
        \draw (-\x*\doubledx-0.4,3.3) node {\scriptsize $B_N$};
        }
        \draw[thick, fill=tensorcolor, rounded corners=2pt] (-3.5*\doubledx,1) rectangle (0.5*\doubledx,0);
	    \draw (-1.5*\doubledx,0.5) node {$\ket{\phi_U}$};
    \end{tikzpicture}
    \end{array}
        ,
    \end{align}
where $B_k$ corresponds to the output, $A_k$ to the input of the unitary, and each $V_k$ is an isometry $V_k^\dagger V_k = \1$. Observe that, due to unitarity, $\ket{\phi_U}$ can be transformed into a maximally entangled state (i.e., $\ket{U}_{AB}$ in the $AB$ bipartition) by the action of local isometries. States with this property (when $V_k$ is non-trivial) have a particular structure and are called LME~\cite{kruszynska2009local}.


Clearly every $\ket{U}_{AB}$ can be expressed in the form of \cref{eq:compressed_CJ} by setting $V_k = \1_{d^2}$. However, the interesting case is when a nontrivial compression is possible, i.e., $\ket{\phi_U}$ has lower dimension that $\ket{U}_{AB}$. Although this may not always be feasible, the following example shows a simple class of unitaries that admit a compression.
\begin{example}[Phase unitaries] \label{ex:phase_unitaries}
    Consider an $N$-qubit unitary diagonal in the product basis
    \begin{align} \label{eq:phase_unitary_qubit}
        U_{\rm phase} = \sum_{i_k = 1}^d e^{i \theta_{i_1 \dots i_N}} \ket{i_1 \dots i_N} \bra{i_1 \dots i_N}.
    \end{align}
    Clearly we can also express
    \begin{align}\label{eq:phase_unitary_qubit_vectorized}
        \ket{U_{\rm phase}}_{AB}  =
    \begin{array}{c}
    \begin{tikzpicture}[scale=.6]
		      \foreach \x in {0,...,0}{
        \draw[very thick] (-\x*\doubledx,0) -- (-\x*\doubledx,1.5);
        \draw[very thick] (-\x*\doubledx,1.5) -- (-\x*\doubledx-0.3,1.5) -- (-\x*\doubledx-0.3,2);
        \draw[very thick] (-\x*\doubledx,1.5) -- (-\x*\doubledx+0.3,1.5) -- (-\x*\doubledx+0.3,2);
		\fill[color=black, thick] (-\x*\doubledx,1.5) circle (0.15);
        \draw (-\x*\doubledx+0.4,2.3) node {\scriptsize $A_1$};
        \draw (-\x*\doubledx-0.4,2.3) node {\scriptsize $B_1$};
        }
		      \foreach \x in {1,...,1}{
        \draw[very thick] (-\x*\doubledx,0) -- (-\x*\doubledx,1.5);
        \draw[very thick] (-\x*\doubledx,1.5) -- (-\x*\doubledx-0.3,1.5) -- (-\x*\doubledx-0.3,2);
        \draw[very thick] (-\x*\doubledx,1.5) -- (-\x*\doubledx+0.3,1.5) -- (-\x*\doubledx+0.3,2);
		\fill[color=black, thick] (-\x*\doubledx,1.5) circle (0.15);
        \draw (-\x*\doubledx+0.4,2.3) node {\scriptsize $A_2$};
        \draw (-\x*\doubledx-0.4,2.3) node {\scriptsize $B_2$};
        }
		      \foreach \x in {2,...,2}{
        \draw[very thick, dotted] (-\x*\doubledx-0.5,1.5) -- (-\x*\doubledx+0.5,1.5);
        }
		      \foreach \x in {3,...,3}{
        \draw[very thick] (-\x*\doubledx,0) -- (-\x*\doubledx,1.5);
        \draw[very thick] (-\x*\doubledx,1.5) -- (-\x*\doubledx-0.3,1.5) -- (-\x*\doubledx-0.3,2);
        \draw[very thick] (-\x*\doubledx,1.5) -- (-\x*\doubledx+0.3,1.5) -- (-\x*\doubledx+0.3,2);
		\fill[color=black, thick] (-\x*\doubledx,1.5) circle (0.15);
        \draw (-\x*\doubledx+0.4,2.3) node {\scriptsize $A_N$};
        \draw (-\x*\doubledx-0.4,2.3) node {\scriptsize $B_N$};
        }
        \draw[thick, fill=tensorcolor, rounded corners=2pt] (-3.5*\doubledx,1) rectangle (0.5*\doubledx,0);
	    \draw (-1.5*\doubledx,0.5) node {$\ket{\psi_{\rm{phase}}}$};
    \end{tikzpicture}
    \end{array}
    \end{align}
    where the LME state is
\begin{align} \label{eq:LME_phase}
    \ket{\psi_{\rm{phase}}} = \sum_{i_k = 0}^1 e^{i \theta_{i_1 \dots i_N}} \ket{i_1 \dots i_N}
\end{align}   
    and each tensor on top represents a Kronecker delta, playing the role of the entangling isometries $V: \ket{i} \mapsto \ket{ii}$.
\end{example}

\subsection{Classification of qubit LME states and their unitaries}

Here we investigate the structure of unitaries whose LME state $\ket{\phi_U}$ admits a nontrivial compression. Our main result establishes the generality of phase unitaries in \cref{ex:phase_unitaries} (proof can be found in \cref{sec:app:proofs_lme}).

\begin{restatable}[]{proposition}{phaseunitaries} \label{prop:lme_unitaries}
Every $N$-qubit unitary admitting a LME state $\ket{\phi_U} \in (\mathbb C ^{2})^{\otimes N}$ is LU-equivalent to a phase unitary [\cref{eq:phase_unitary_qubit}].
\end{restatable}

\noindent Here LU-equivalence means up to product unitaries $U \stackrel{LU}{=} (\bigotimes_k V_k) U (\bigotimes_k W_k)$. As a corollary, this result also extends the characterization of multi-qubit LME states of Ref.~\cite{kruszynska2009local} (see \cref{sec:app:proofs_lme}).

The preceding result does not rely on any tensor-network assumptions. Nevertheless, every MPU can be efficiently brought into the form \cref{eq:compressed_CJ} by a singular-value decomposition over each tensor. The corresponding LME state $\ket{\phi_U}$ is then a MPS with the same bond dimension. If the MPU is in the canonical form [\cref{eq:gauge_nonuniform}], so is the LME-MPS. In that sense, the classification of nonuniform MPU is closely related to the characterization of LME-MPS.

\begin{example}[Phase MPU and finite automata] \label{ex:cellular_automata}
    Consider a multi-qubit ($d=2$) MPU admitting a LME state of uniform physical dimension $d'=2$. According to \cref{prop:lme_unitaries}, this class corresponds to phase unitaries (up to LU), such that $\ket{\psi_{\rm {phase}}}$ is a MPS [cf.~\cref{ex:phase_unitaries}].
    
    A subclass of these MPS can be produced by tensors of the form
    \begin{align} \label{eq:tensor_automaton}
        \begin{array}{c}
            \begin{tikzpicture}[scale=0.6,baseline={([yshift=-8ex] current bounding box.center)}]
		\draw[very thick] (0,0) -- (0,1);
		\draw[very thick] (-1,0) -- (1,0);
        \draw[ thick, fill=tensorcolor, rounded corners=2pt] (-0.5,-0.5) rectangle (0.5,0.5);
		\draw (0,0) node {\small $A_k$};
		\draw (-1.4,0) node {$i$};
		\draw (1.4,0) node {$l$};
		\draw (0,1.4) node {$j$};
            \end{tikzpicture}
        \end{array}
        \!\!\!= \delta_{l,f^{(k)}_{ij}} \exp (i \theta^{(k)}_{ij} ) \;,
    \end{align}
where $f^{(k)}_{ij} \in \{0, \dots, D_{k-1} -1 \}$ and $\theta^{(k)}_{ij}$ are arbitrary phases. The tensor can be understood as encoding a deterministic weighted finite automaton~\cite{droste2009handbook}. Its weights correspond to the complex phases, while its memory is bounded by the bond dimension $D$ and the transitions are specified by the function $f$.
    
For instance, this subclass includes $\ket{\psi_{\rm phase}}$ corresponding to the multi-control $Z$-gate of \cref{ex:multi_control_z}. For this case $D=2$, $i,j \in \{  0,1 \}$ and  $f_{ij} = i \cdot j$, while there is a single nontrivial $\theta^{(1)}_{1,1} = \pi$. The leftmost $A_N$ can be taken to be the identity tensor.
\end{example}

Another reason the subclass of \cref{eq:tensor_automaton} is special is that the corresponding MPU admits a direct quantum circuit implementation with the aid of ancillas.


\begin{example}[Circuit implementation]
    Let $U_{\rm phase}$ be a unitary as in \cref{eq:phase_unitary_qubit_vectorized} where $\ket{\psi_{\rm phase}}$ is a LME-MPS of the form \eqref{eq:tensor_automaton}. Then $U_{\rm phase}$ is a MPU and can be implemented using $N-2$ $D$-level ancillas initialized in $\ket{0}$ as
    \begin{align}
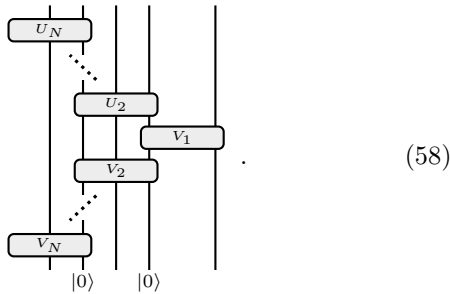
                         \begin{array}{c}
        \begin{tikzpicture}[scale=.6]
		      \foreach \x in {1,...,3}{
      \draw [thick] (-\doubledx*\x,-2.4) -- (-\doubledx*\x,4);
        }
      \draw [thick] (-\doubledx*2.5,-2.4) -- (-\doubledx*2.5,4);
     \draw [thick] (-\doubledx*3.5,-2.4) -- (-\doubledx*3.5,4);
	        \draw (-\doubledx*2,-2.7) node {\scriptsize $\ket{0}$};
	        \draw (-\doubledx*3,-2.7) node {\scriptsize $\ket{0}$};
		\foreach \x in {1,...,1}{
        \gate{(-\doubledx*\x-\doubledx*0.5,\doubledx/2*\x)}{\tiny $V_\x$};          
        }
		\foreach \x in {2,...,2}{
        \gate{(-\doubledx*\x-\doubledx*0.5,\doubledx/2*\x)}{\tiny $U_\x$};          
        }
		\fill[fill=white] (-3.1*\doubledx,2.8) rectangle (-2.9*\doubledx,2.2);
        \draw [very thick, dotted] (-2.8*\doubledx,2.2) to (-3.2*\doubledx,2.8);
        \begin{scope}[shift={(0,1)}]
		\foreach \x in {3,...,3}{
        \gate{(-\doubledx*\x-\doubledx*0.5,\doubledx/2*\x)}{\tiny $U_{N}$};          
        }
        \end{scope}
    \begin{scope}[shift={(0,1.6)}]
		\foreach \x in {2,...,2}{
        \gate{(-\doubledx*\x-\doubledx*0.5,-\doubledx/2*\x)}{\tiny $V_\x$};          
        }
		\fill[fill=white] (-3.1*\doubledx,-2.8) rectangle (-2.9*\doubledx,-2.2);
        \draw [very thick, dotted] (-2.8*\doubledx,-2.2) to (-3.2*\doubledx,-2.8);
        \begin{scope}[shift={(0,-1)}]
		\foreach \x in {3,...,3}{
        \gate{(-\doubledx*\x-\doubledx*0.5,-\doubledx/2*\x)}{\tiny $V_{N}$};          
        }
        \end{scope}
    \end{scope}
        \end{tikzpicture}
                \end{array}
        \;.
    \end{align}
The $k$-th unitary gate ($N-1 \ge k \ge 2$) acts over $\mathbb C ^{D_{k}} \otimes \mathbb C ^{d} \otimes \mathbb C ^{D_{k-1}}$ as
\begin{subequations}
    \begin{align}
        V_k \ket{ij}\ket{l} &= \exp (i \theta^{(k)}_{ij} )\ket{ij}\ket{l + f_{ij}^{(k)}} \\
        U_k \ket{ij}\ket{l} &= \ket{ij}\ket{l - f_{ij}^{(k)}} 
    \end{align}
\end{subequations}
where addition and subtraction are $({\rm mod} \ D_{k-1})$. The boundary unitaries for $k = N$ are defined similarly but for a single ancilla instead of two, while $V_1 \ket{ij} = \exp (i \theta^{(1)}_{ij} ) \ket{ij}$.

It is direct to check that the above circuit implements the desired phase unitary. The key point is that, since $V_k$, $U_k$ are control unitaries and $f^{(k)}_{ij}$ has a definite value, the ancillas return to $\ket{0}$ in the end of the circuit for every input state.
\end{example}

\section{Summary and Outlook} \label{sec:outlook}

In this paper, we make some initial steps towards developing a theory of MPU that goes beyond QCA, i.e., beyond MPU formed by a single repeated tensor and periodic boundary conditions. There are many possibilities in that direction. Perhaps the simplest is by adding a boundary condition over a bulk that consists of a single repeated tensor. We show that this can be consistent with translation-invariance of the resulting unitary operator. Nevertheless, the freedom of the boundary condition is enough to maximally violate the QCA property. We identify permutation-invariant and translation-invariant families admitting a block-diagonal decomposition while not being a QCA. Motivated by this, we systematically investigate MPU with a block-diagonal structure, which we call semi-simple. We show that the semi-simple property is incompatible with QCA, unless the latter are trivial product unitaries.

Departing from the case of uniform bulk, we consider the general case of MPU formed by arbitrary tensors. We formulate unitarity in the auxiliary space and prove that this class cannot always be written as sequential circuits, even if two sequential layers are allowed. This is in contrast with MPS, where a single layer always suffices. We characterize MPU that can be locally isometrically compressed to $d=2$ showing that they are local-unitary equivalent to unitaries with product eigenvectors.

Many interesting questions remain open. First of all, one could consider generalizing to matrix-product isometries. Already for the unitary case, an exhaustive characterization or parametrization of MPU is missing. Moreover, it is not clear what the circuit complexity of a general MPU is, or even if it is polynomial in the number of sites. For MPS, their $O(N)$ complexity follows easily from the canonical form. In that respect, it is also unclear how important is the role of ancillary qubits. In particular, we have seen that ancillas can aid the implementation of certain MPU examples (like the multi-control $Z$-gate). However, in the case of QCA, access to ancillas trivializes all of them, i.e., reduces them to a finite-depth local quantum circuit. For uniform MPU with boundary, we focus here on the structure of semi-simple ones, but in general a classification of their correlations, as well as a canonical form, are missing.

\section*{Acknowledgements}

GS is grateful to Marta Florido-Llin\'{a}s for many insightful discussions. RT acknowledges startup funding from University of Washington. This research is part of the Munich Quantum Valley (MQV), which is supported by the Bavarian state government with funds from the Hightech Agenda Bayern Plus. The work has been financially supported by Universidad Complutense de Madrid (grant FEI-EU-22-06), by the Spanish Ministry of Science and Innovation MCIN/AEI/10.13039/501100011033 (``Severo Ochoa Programme for Centres of Excellence in R\&D'' CEX2019-000904-S and grant PID2020-113523GB-I00), by the Ministry for Digital Transformation and of Civil Service of the Spanish Government through the QUANTUM ENIA project call – Quantum Spain project, and by the European Union through the Recovery, Transformation and Resilience Plan – NextGenerationEU within the framework of the Digital Spain 2026 Agenda.
This research was supported in part by Perimeter Institute for Theoretical Physics. Research at Perimeter Institute is supported by the Government of Canada through the Department of Innovation, Science, and Economic Development, and by the Province of Ontario through the Ministry of Colleges and Universities.

\bibliographystyle{quantum}
\bibliography{my_refs}

\appendix

\setcounter{equation}{0}
\setcounter{figure}{0}
\setcounter{table}{0}
\makeatletter
\renewcommand{\theequation}{S\arabic{equation}}
\renewcommand{\thefigure}{S\arabic{figure}}

\section{Proofs of \cref{sec:inhomo_structure}} \label{sec:app:proofs_inhomo_canonical}


%

\subsection*{Proof of \cref{prop:doublestaircase}}


\doublestaircase*
\begin{proof}
In our setup we have $N+1$ qubits, which we enumerate from left to right. The main idea is to consider the operator Schmidt decomposition over the bipartition $(1 \dots N)-(N+1)$. If \cref{eq:app:left-right-staircase} holds, then a necessary condition is that the Schmidt spectrum (i.e., singular values) over the above bipartition are the same. We will show that this is not always possible.

For this, the key observation is that the unitary gates $U_2 \dots U_N$ and $V_2 \dots V_N$ do not affect the Schmidt spectrum, and can thus be replaced by identity. Thus it suffices to show that the Schmidt spectra of the following operators
    \begin{align} \label{eq:app:left-right-staircase_schmidt}
    s \Big[
        \begin{array}{c}
        \begin{tikzpicture}[scale=0.6]
		      \foreach \x in {1,...,4}{
      \draw [thick] (\doubledx*\x,-2.4) -- (\doubledx*\x,4);
        }
		\foreach \x in {1,...,2}{
        \gate{(\doubledx*\x+\doubledx*0.5,\doubledx/2*\x)}{};          
        }
		\fill[fill=white] (3.1*\doubledx,2.8) rectangle (2.9*\doubledx,2.2);
        \draw [very thick, dotted] (2.8*\doubledx,2.2) to (3.2*\doubledx,2.8);
        \begin{scope}[shift={(0,1)}]
		\foreach \x in {3,...,3}{
        \gate{(\doubledx*\x+\doubledx*0.5,\doubledx/2*\x)}{};          
        }
        \end{scope}
    \begin{scope}[shift={(0,1.6)}]
		\foreach \x in {2,...,2}{
        \gate{(\doubledx*\x+\doubledx*0.5,-\doubledx/2*\x)}{}; 
        }
		\fill[fill=white] (3.1*\doubledx,-2.8) rectangle (2.9*\doubledx,-2.2);
        \draw [very thick, dotted] (2.8*\doubledx,-2.2) to (3.2*\doubledx,-2.8);
        \begin{scope}[shift={(0,-1)}]
		\foreach \x in {3,...,3}{
        \gate{(\doubledx*\x+\doubledx*0.5,-\doubledx/2*\x)}{};          
        }
        \end{scope}
    \end{scope}
      \draw [very thick,dashed] (3.5*\doubledx,-2.8) -- (3.5*\doubledx,4.4);
        \end{tikzpicture}
                \end{array}
        \Big]
            \ne
        s \Big[
                 \begin{array}{c}
        \begin{tikzpicture}[scale=0.6]
		      \foreach \x in {1,...,4}{
      \draw [thick] (-\doubledx*\x,-2.4) -- (-\doubledx*\x,4);
        }
		\foreach \x in {1,...,1}{
        \gate{(-\doubledx*\x-\doubledx*0.5,\doubledx/2*\x)}{};          
        }
      \draw [very thick,dashed] (-1.5*\doubledx,-2.8) -- (-1.5*\doubledx,4.4);
        \draw [very thick, dotted] (-3.2*\doubledx,.75) to (-3.8*\doubledx,.75);
        \end{tikzpicture}
                \end{array}
        \Big]
\end{align}
are distinct (the dashed line denotes the bipartition).

We now use the standard decomposition of the 2-qubit gate $U_1$, which parameterizes its nonlocal part via three angle variables~\cite{kraus2001optimal}. From this one can obtain the explicit form of the Schmidt coefficients (see Ref.~\cite{dur2002optimal}). Using this expression, it follows that the sum of the squares of the Schmidt coefficients is constant, which we can normalize to one. Crucially, however, the corresponding probability vector does not span the entire 3-simplex when the angle variables are varied.

It thus remains to show that there is a choice of gates for the lhs of \cref{eq:app:left-right-staircase_schmidt} which can yield a Schmidt spectrum that does not belong in the aforementioned subregion of the 3-simplex. This is indeed possible, already by a suitable choice of $\widetilde{U}_N, \widetilde{U}_{N-1}, \widetilde{V}_N$ and the rest of the gates acting as identity (thus $N=2$ already suffices).

For instance, taking
\begin{subequations}
\begin{gather}
    \widetilde{U}_{N} = \exp \left[ i\left(\frac{\pi}{4} \sigma_X\otimes \sigma_X + \frac{\pi}{4} \sigma_Y \otimes \sigma_Y + \theta_U \sigma_Z \otimes \sigma_Z\right) \right]\\
    \widetilde{U}_{N-1} = \mathrm{CNOT} \\
    \widetilde{V}_{N} = \exp \left[ i\left(\frac{\pi}{4} \sigma_X\otimes \sigma_X + \frac{\pi}{4} \sigma_Y \otimes \sigma_Y + \theta_V \sigma_Z \otimes \sigma_Z\right)\right]
\end{gather}
\end{subequations}
with the rest of the gates identity gives a Schmidt spectrum 
\begin{align}
    s = \{ 2 |\cos \theta_+|, 2|\sin \theta_+|, 2|\cos \theta_-|, 2|\sin \theta_-|  \}
\end{align}
where $ \theta_{\pm} = \theta_U \pm \theta_V$. Using the expressions from Ref.~\cite{dur2002optimal}, it is easy to check that the above range is outside of the Schmidt spectrum of a single 2-qubit gate.
\end{proof}

\section{Proofs of \cref{sec:homo}} \label{sec:app:proofs_homo}

\subsection*{Proof of \cref{prop:structure_semisimple}}

Here prove \cref{prop:app:structure_semisimple_extended} which is a refinement of \cref{prop:structure_semisimple}. For this we will need the following simple Lemma, which has been noted in \cite{degroot2022symmetry}, but we repeat its proof here for convenience.

\begin{alemma}\label{app:lemma:vandermonde}
    Let $\lambda_i \in \mathbb C$  with $\lambda_i \ne 0$ and $\lambda_i \ne \lambda_j$ for all $i,j=1,\dots,N$. Then for any integer $k\ge 0$
    \begin{align} \label{eq:vandermonde}
        \sum_{i=1}^{N} c_i \lambda_i^n = 0 \; \forall n = k,\dots,k+N-1 \quad 
        \Longrightarrow \quad c_i = 0 \;  \forall i \,.
    \end{align}
\end{alemma}
\begin{proof}
    The implication is true if and only if the vectors $\{ v_n\}_{n=k}^{k+N-1}$ are linearly independent, where $v_n = (\lambda_1^n,\dots,\lambda_N^n)$. This corresponds to the determinant of a Vandermonde matrix, which is indeed nonvanishing given the assumptions, as is can be checked from its explicit expression~\cite{horn2012matrix}.
\end{proof}

\begin{aproposition}[Structure of semi-simple MPU]\label{prop:app:structure_semisimple_extended}
    After blocking a finite number of times, every semi-simple MPU satisfies
    \begin{align}
        \mathcal S E = E \mathcal S \subseteq \mathcal S ^2 \,.
    \end{align}
    In particular, it admits a representation such that, up to a similarity transformation over $\mathbb C^D \otimes \mathbb C^D$,
    \begin{subequations}
        \begin{alignat}{6}
            E &= \1_q && \oplus Q''_E  &&\oplus Q'_E &&\oplus 0 &&\oplus J_E &&{} \\
            \mathcal S &= 0_q && \oplus 0 &&\oplus Q'_S &&\oplus 0 &&\oplus J_S &&{} \\
            B &= B_{\1} && \oplus B''_Q &&\oplus  B'_Q &&\oplus J_B &&\oplus 0 && + \text{off-diagonal.}
        \end{alignat}
%
%
%
    \end{subequations}
Above $Q'_E \in Q'_S$, where
\begin{align}
        Q'_S &= \bigoplus_j \bigoplus_k  \lambda_{jk} \1_{r_{jk}} \otimes \mathbb M_{d_j} \\
        B'_Q &= \bigoplus_j \bigoplus_{k} \bigoplus_{l=1}^{r_{jk}} B_{jkl}
\end{align}
    satisfying:
    \begin{enumerate}[(i)]
        \item $\lambda_{jk} \ne 0$, $\lambda_{jk} \ne \lambda_{jk'}$ if $k \ne k'$, $r_{jk} \ge 2$, $B_{jkl} \ne 0$, $\sum_{l} B_{jkl} = 0$.
        \item As in (i) for $Q''_E$, $B''_Q$ but with $d_j = 1$.
        \item $q \ge 1$ and $\Tr B_\1 = 1$.
    \end{enumerate}
\end{aproposition}
\begin{proof}
    As explained in the main text, assuming that $\mathcal A = \bigoplus_i A_i$ where $\mathcal A_i$ are irreducible implies that blocking a finite number of times leads to
\begin{align}
     \mathcal A = \bigoplus_j \diag (\lambda_{j1},\dots, \lambda_{jm_j}) \otimes \mathbb M_{d_j}
\end{align}
up to a similarity transformation in $\mathbb C^D \otimes \mathbb C^D$.

Without loss of generality, we separate 
\begin{align}
     \mathcal A = \bigoplus_j \diag (\lambda_{j1},\dots, \lambda_{jm_j}) \otimes \mathbb M_{d_j} \bigoplus 0
\end{align}
    such that we can now assume that $\lambda_{jk} \ne 0$.
In the same basis,
\begin{align}
    E &= \bigoplus_j \diag (\lambda_{j1},\dots, \lambda_{j m'_j}) \otimes E_{j} \bigoplus 0 \\
     \mathcal S &= \bigoplus_j \diag (\lambda_{j1},\dots, \lambda_{j m'_j}) \otimes R_{j} \bigoplus 0
\end{align}
where the $\lambda$'s are the same as above but $R_j$ are, for the moment, arbitrary subspaces. There are four possibilities for each $R_j$:
\begin{enumerate}[(a)]
    \item $R_j = 0$
    \item $R_j = \mathbb M_{d_j}$
    \item $R_j \ne 0$ and $R_j \subset \mathbb M_{d_j}$ irreducible
    \item $R_j \ne 0$ and $R_j \subset \mathbb M_{d_j}$ reducible
\end{enumerate}
We will now show that by blocking further, we can reduce to (a) or (b).

Consider possibility (c). Then the span either corresponds to a normal tensor (c1) or has periodic subspaces (c2). For (c1) we can block further to reduce it to (b). Case (c2) is impossible. This is since $\dim R_j = d_j - 1$ or $\dim R_j = d_j$ because $\mathcal A$ and $\mathcal S$ differ by a single element in the generators of the span. However, the smallest number of periodic subspaces is 2, and thus (c2) would imply $\dim R_j \le d_j - 2$.

Considering (d), the only reducible span that may become a full algebra with the addition of a single element is of the form
$R_j = \begin{pmatrix}
    * & * \\
    0 & *
\end{pmatrix}$ for $d_j = 2$. However, then $E_j$ has support on the lower left corner, and thus blocking twice leads to (b).

Henceforth we assume that we have blocked such that either (a) or (b) holds. Notice that blocking redefines $\lambda_{jk}$ and $E$, but in what follows we keep the same notation.

Suppose there are $q'$ blocks corresponding to an index $j$ such that (a) above holds. Since $\lambda_{jk} \ne 0$, necessarily $d_j = 1$ and $E_j \ne 0$ for all such $j$. We now separate these blocks in our notation (arranging them to the leftmost of the direct sum) since they will play a role later on:
\begin{subequations} \label{app:subeq:ideal}
\begin{align}
    E &= T \bigoplus_j \diag (\lambda_{j1},\dots, \lambda_{j m''_j}) \otimes E_{j} \bigoplus 0 \\
    \mathcal S &= 0 \bigoplus_j \diag (\lambda_{j 1},\dots, \lambda_{j m''_j}) \otimes \mathbb M_{d_j} \bigoplus 0
\end{align}
\end{subequations}
\noindent where $T = \diag(t_1,t_2,\dots)$ with $t_i \ne 0$ and we have again relabeled $\lambda_{jk}$.

Now let us examine the boundary conditions. Expressed in the same basis as $E$ and $\mathcal S$ they may fail to be block diagonal. Nevertheless, the conditions of \cref{lem:unitarity_bond_S_homo} exhibit a trace in which the off-diagonal blocks of $B$ do not contribute, thus we will only write for convenience the block-diagonal part. In particular, it might happen that some blocks of $B$ are identically vanishing. Separating these blocks (rightmost in the direct sum) we can write
        \begin{alignat}{4}
            E &= T  &&\oplus Q'_E &&\oplus 0 &&\oplus J_E  \\
            \mathcal S &= 0_q &&\oplus Q'_S &&\oplus 0 &&\oplus J_S \\
            B &= B_T &&\oplus  B'_Q &&\oplus J_B &&\oplus 0 
        \end{alignat}
where we have redefined $T$ and once again relabeled $\lambda_{jk}$, obtaining
\begin{subequations} \label{app:subeq:QEQS}
\begin{align}
    Q'_E &= \bigoplus_j \diag (\lambda_{j1},\dots, \lambda_{j m'''_j}) \otimes E_{j}  \\
    Q'_S &= \bigoplus_j \diag (\lambda_{j1},\dots, \lambda_{j m'''_j}) \otimes \mathbb M_{d_j} \,.
\end{align}
\end{subequations}

We are now ready to impose the unitarity equations of \cref{lem:unitarity_bond_S_homo}. From \cref{app:subeq:ideal} we have $\mathcal S E = E \mathcal S \subseteq \mathcal S ^2$. Thus there are only two independent equations:
\begin{align}
    \Tr \left(B \mathcal S ^n\right) = 0 \quad \forall n \label{eq:app:Sn_zero}\\
    \Tr \left(B E ^n\right) = 1 \quad \forall n \label{eq:app:En_one}
\end{align}
\noindent Notice that the blocks $J_E, J_S, J_B$ do not contribute to above trace.

We begin with \cref{eq:app:Sn_zero}. Denoting
\begin{align}
    B'_Q = \bigoplus_j \bigoplus_{k=1}^{m'''_j} B_{jk} \,,
\end{align}
it implies that
\begin{align}
    \sum_{j,k} \Tr( B_{jk} M_j ) \lambda_{jk}^n = 0 \quad \forall M_{j} \in \mathbb M_{d_j}, \,\forall n \,.
\end{align}
Taking into account degeneracies in $\lambda_{jk}$, we can now apply \cref{app:lemma:vandermonde}. It gives
\begin{align}
    \sum_{j,k} \Tr( B_{jk} M_j ) = 0 \quad \forall M_{j} \in \mathbb M_{d_j} 
\end{align}
where the sum above runs over pairs of indices $(j,k)$ such that $\lambda_{jk}$ remains unchanged. Since by assumption $B_{jk} \ne 0$, we get
\begin{align}
        \sum_{k' \in \Lambda (\lambda_{jk})} B_{jk'}  = 0 \quad \forall j \,\; \forall \lambda_{jk} \,,
\end{align}
where $\Lambda(\lambda_{jk})$ is the set of indices defined by $k' \in \Lambda(\lambda_{jk}) \Leftrightarrow \lambda_{jk} = \lambda_{jk'}$. We have thus obtained that, for any given $j$, the different $\lambda_{jk}$ that appear in the decomposition \cref{app:subeq:QEQS} are necessarily at least twofold degenerate, while the corresponding blocks $B_{jk}$ of the boundary have a vanishing sum. Relabeling yet again to make the degeneracies explicit, we obtain the form
\begin{align}
        Q'_S = \bigoplus_j \bigoplus_k  \lambda_{jk} \1_{r_{jk}} \otimes \mathbb M_{d_j} 
\end{align}

Finally, we impose \cref{eq:app:En_one}. Since $Q'_E \in Q'_S$, we immediately have
\begin{align}
    \Tr (B_0 T^n) = 1 \quad \forall n \,.
\end{align}
Since both are diagonal matrices with nonzero elements in the diagonal, \cref{app:lemma:vandermonde} implies that (i) 1 belongs in the diagonal of $T$ with multiplicity $q\ge1$, (ii) the rest of the eigenvalues of $T$ are at least twofold degenerate, as for $Q'_S$. We can split these two contributions and express $T = \1_q \oplus Q''_E$. This leads to
        \begin{alignat}{5}
            E &= \1_q && \oplus Q''_E  &&\oplus Q'_E &&\oplus 0 &&\oplus J_E  \\
            \mathcal S &= 0_q && \oplus 0 &&\oplus Q'_S &&\oplus 0 &&\oplus J_S \\
            B &= B_{\1} && \oplus B''_Q &&\oplus  B'_Q &&\oplus J_B &&\oplus 0 \,.
        \end{alignat}
with the constraints as announced.
\end{proof}

\section{Proofs of \cref{sec:lme}} \label{sec:app:proofs_lme}

Let us first recall the definition of a LME-state, first introduced in~\cite{kruszynska2009local}.
\begin{adefinition}[Ref.~\cite{kruszynska2009local}]
    A multipartite state $\ket{\psi_{\rm{LME}}}$ is LME if there exist local isometries $V_k: \mathbb C^{d_k'} \to \mathbb C^{d_k} \otimes \mathbb C^{d_k}$ with $V_k^\dagger V_k = \1_{d_k'}$ such that
    \begin{align} \label{eq:lme_def}
        \ket{\widetilde{\psi}_{\rm{LME}}}_{AB} = \bigotimes_k V_k \ket{\psi_{\rm{LME}}} =
    \begin{array}{c}
    \begin{tikzpicture}[scale=.6]
		      \foreach \x in {0,...,0}{
        \draw[very thick] (-\x*\doubledx,0) -- (-\x*\doubledx,1.5);
        \draw[very thick] (-\x*\doubledx+0.3,2.5) -- (-\x*\doubledx+0.3,3);
        \draw[very thick] (-\x*\doubledx-0.3,2.5) -- (-\x*\doubledx-0.3,3);
        \draw[thick, fill=whitetensorcolor, rounded corners=2pt] (-\x*\doubledx+.5,1.5+1) rectangle (-\x*\doubledx-.5,1.5); 
        \draw (-\x*\doubledx,2) node {\small $V_1$};
        \draw (-\x*\doubledx+0.4,3.3) node {\scriptsize $A_1$};
        \draw (-\x*\doubledx-0.4,3.3) node {\scriptsize $B_1$};
        }
		      \foreach \x in {1,...,1}{
        \draw[very thick] (-\x*\doubledx,0) -- (-\x*\doubledx,1.5);
        \draw[very thick] (-\x*\doubledx+0.3,2.5) -- (-\x*\doubledx+0.3,3);
        \draw[very thick] (-\x*\doubledx-0.3,2.5) -- (-\x*\doubledx-0.3,3);
        \draw[thick, fill=whitetensorcolor, rounded corners=2pt] (-\x*\doubledx+.5,1.5+1) rectangle (-\x*\doubledx-.5,1.5); 
        \draw (-\x*\doubledx,2) node {\small $V_2$};
        \draw (-\x*\doubledx+0.4,3.3) node {\scriptsize $A_2$};
        \draw (-\x*\doubledx-0.4,3.3) node {\scriptsize $B_2$};
        }
		      \foreach \x in {2,...,2}{
        \draw[very thick, dotted] (-\x*\doubledx-0.5,2) -- (-\x*\doubledx+0.5,2);
        }
		      \foreach \x in {3,...,3}{
        \draw[very thick] (-\x*\doubledx,0) -- (-\x*\doubledx,1.5);
        \draw[very thick] (-\x*\doubledx+0.3,2.5) -- (-\x*\doubledx+0.3,3);
        \draw[very thick] (-\x*\doubledx-0.3,2.5) -- (-\x*\doubledx-0.3,3);
        \draw[thick, fill=whitetensorcolor, rounded corners=2pt] (-\x*\doubledx+.5,1.5+1) rectangle (-\x*\doubledx-.5,1.5); 
        \draw (-\x*\doubledx,2) node {\small $V_N$};
        \draw (-\x*\doubledx+0.4,3.3) node {\scriptsize $A_N$};
        \draw (-\x*\doubledx-0.4,3.3) node {\scriptsize $B_N$};
        }
        \draw[thick, fill=tensorcolor, rounded corners=2pt] (-3.5*\doubledx,1) rectangle (0.5*\doubledx,0);
	    \draw (-1.5*\doubledx,0.5) node {$\ket{\psi_{\rm{LME}}}$};
    \end{tikzpicture}
    \end{array}
        .
    \end{align}
    is maximally entangled in the $AB$ bipartition. If $d_k = d$ $\forall k$ we say $\ket{\psi_{\rm{LME}}}$ is a $(d,d')$-LME.
\end{adefinition}

LME-states can thus be transformed into maximally entangled states by first adding local auxiliary degrees of freedom and then applying joint local unitaries to each system-ancilla pair. For instance, all product states have this property. In general, however, it is a nontrivial constraint.

Ref.~\cite{kruszynska2009local} characterized all $(2,2)$-LME states under the constraint that all $V_k$ are control isometries, i.e., arise from 2-qubit control unitaries by fixing an input state.
\begin{aproposition}[Ref.~\cite{kruszynska2009local}] \label{lem:barbara}
    Up to LU-transformations, all $(2,2)$-LME states for which $V_k$ can be chosen to be control isometries have the form
\begin{align} \label{eq:LME_barbara}
    \ket{\psi_{\rm{phase}}} = \frac{1}{\sqrt{2^N}} \sum_{i_k = 0}^1 e^{i \theta_{i_1 \dots i_N}} \ket{i_1 \dots i_N}\;.
\end{align}
    Conversely, every such state is LME and $V_k$ can be chosen to be control isometries.
\end{aproposition}
\noindent We later generalize this result in \cref{cor:app:strong_lme}.

Before proving \cref{prop:lme_unitaries} we will reformulate the LME property in the language of quantum channels. Due to the constrained dimensionality of the LME state, proving the result will be related to properties of quantum channels acting on a single qubit, the structure of which has been analyzed in detail in Ref.~\cite{ruskai2002analysis}.

\begin{alemma} \label{app:lem:lme_channels}
    A state $\ket{\psi_{\rm {LME}}}$ over $(\mathbb C^{d'})^{\otimes N}$ is a $(d,d')$-LME if and only if there exists local quantum channels $\mathcal E_k$ of Kraus rank $r_K \le d$ such that
    \begin{align}
        \bigotimes_{k=1}^N \mathcal E_k (\ket{\psi_{\rm {LME}}} \bra{\psi_{\rm {LME}}}) = \frac{\1}{d^N}
    \end{align}
\end{alemma}
\begin{proof}
    From the definition of the LME property, maximal entanglement of $\ket{\widetilde{\psi}_{\rm {LME}}}_{AB}$ [cf. \cref{eq:lme_def}] is equivalent to 
    \begin{align} \label{eq:app:channel_lme}
        \Tr_B (\ket{\widetilde{\psi}_{\rm {LME}}} \bra{\widetilde{\psi}_{\rm {LME}}} ) = \frac{\1}{d^N} \;.
    \end{align}
    This last property amounts to the action of the channels $\mathcal E_k$ on $\ket{\psi_{\rm{LME}}} \bra{\psi_{\rm{LME}}}$ with Kraus operators $K^{(k)}_i = \prescript{}{B}{\bra{i}} V_{k}$. Since $i = 1 , \dots, d$ the Kraus rank is at most $d$. This shows the existence of the channels. The converse follows by defining the isometries $V_k$ through the Stinespring isometry corresponding to $\mathcal E_k$. 
\end{proof}

We will soon need this technical lemma.
\begin{alemma} \label{app:lem:ruskai}
    If a quantum channel $\mathcal E$ over a single qubit has $(i)$ Kraus rank $r_K \le 2$, $(ii)$ $\det \mathcal E = 0$, and $(iii)$ $\mathcal E (\rho) = \1/2$ for some state $\rho$, then $\mathcal E$ is dephasing, i.e.,
        \begin{align} \label{eq:app:zdephasing}
            \mathcal E(X) = \frac{1}{2} X + \frac{1}{2} \sigma_z X \sigma_z 
        \end{align}
        up to LU.
\end{alemma}

\noindent LU equivalence here means up to unitaries $V,W$ where $\mathcal E(X) \stackrel{LU}{=} W \mathcal E(V X V^\dagger) W^\dagger$

\begin{proof}
    The result follows from the analysis of Ref.~\cite{ruskai2002analysis}. There, it was shown that for qubit channels with Kraus rank $r_K \le 2$ with $\det \mathcal E = 0$, up to LU,
        \begin{align}
        \mathcal E = 
            \begin{pmatrix}
                1 & 0 & 0 & 0\\
                0 & \cos u & 0 & 0\\
                0 & 0 & 0 & 0\\
                \sin u & 0 & 0 & 0
            \end{pmatrix}
        \text{ or} \; 
        \mathcal E = 
            \begin{pmatrix}
                1 & 0 & 0 & 0\\
                0 & 0 & 0 & 0\\
                0 & 0 & \cos v & 0\\
                \pm \sin u & 0 & 0 & 0
            \end{pmatrix}
        \end{align}
    where $u \in [0,2\pi)$, $v \in [0, \pi)$, and the matrix representation is given in the Pauli basis. From the explicit parametrization above it follows that there is no quantum state such that $\mathcal E (\rho) = \1/2$ unless $\sin u = 0$ and $\sin v = 0$, which yields the desired result.
\end{proof}

We are now ready to prove the main result.

\phaseunitaries*
\begin{proof}
    By \cref{app:lem:lme_channels}, there exist single-qubit quantum channels $\mathcal E_k$ with Kraus rank $r_K \le 2$ such that
    \begin{align} \label{eq:app:lme_channel_qubit}
        \bigotimes_{k} \mathcal E_k  (\ket{\phi_U} \bra{\phi_U}) = \frac{\1}{2^N} \;.
    \end{align}
    Our strategy will be to utilize \cref{app:lem:ruskai}. For that, we begin by showing that condition $(iii)$ of the Lemma is satisfied for all $\mathcal E_k$. Denote
    \begin{align}
        \rho_k = \Tr_{\bar{k}} \ket{\phi_U} \bra{\phi_U}
    \end{align}
    the reduced state over the single site $k$. Taking partial trace over all sites except $k$ in \cref{eq:app:lme_channel_qubit} gives
    \begin{align}
        \mathcal E_k (\rho_k) = \frac{\1}{2}
    \end{align}
    due to trace preservation of the quantum channels.

    Let $A$ denote the set of sites for which $\det \mathcal E_k = 0$ and $B$  its complement. By \cref{app:lem:ruskai}, all $\mathcal E_k$ for $k \in A$ are (up to LU) dephasing channels and the rest invertible maps. Thus
    \begin{align} \label{eq:app:dephasing:lambda_2}
        \bigotimes_{k \in A} \mathcal E_k  (\ket{\phi_U} \bra{\phi_U}) =  \bigotimes_{k \in A} \frac{\1_k}{2} \bigotimes_{k \in B} \mathcal \rho_k \;.
    \end{align}
    Since here we are interested up to LU-equivalence, we can assume without loss of generality that we have $Z$-dephasing, i.e., as in \cref{eq:app:zdephasing}. Then \cref{eq:app:dephasing:lambda_2} takes the form
    \begin{align} \label{eq:app:dephasing:lambda_3}
        \prescript{}{A}{\bra{i}} \phi_U \rangle_{AB} \bra{\phi_U} i \rangle_A = \frac{1}{2^{N_A}} \bigotimes_{k \in B} \rho_k  \quad \forall i\;,
    \end{align}
    where $\ket{i}_A = \ket{i_1 \dots i_{N_A}}_A$ is an element of the product-$Z$ basis over sites in $A$ and $N_A$ is the number of sites in $A$. Thus all $\rho_k$ are pure for $k \in B$ and, up to LU, we can take $\rho_k = \ket{+}\bra{+}$. Plugging this in \cref{eq:app:dephasing:lambda_3} we get
    \begin{align} \label{eq:app:lme_state_general}
        \ket{\phi_U} = \frac{1}{2^{N_A/2}} \left( \sum_{i_k =0}^1 e^{i \theta_{i_1 \dots i_{N_A}}} \ket{i_1 \dots i_{N_A}}      \right) \otimes \ket{+}^{\otimes N_B}  \;.
    \end{align}

    We have therefore shown that, up to LU,
    \begin{align}
        U = \left( \sum_{i_k =0}^1 e^{i \theta_{i_1 \dots i_{N_A}}} \ket{i_1 \dots i_{N_A}}_A \bra{i_1 \dots i_{N_A}}     \right) \otimes \1_B
    \end{align}
    since $U$ factorizes over sites in $B$.
\end{proof}

The proof [\cref{eq:app:lme_state_general}] also generalizes \cref{lem:barbara} by lifting the assumption that $V_k$ are control isometries.

\begin{acorollary} \label{cor:app:strong_lme}
    All $N$-qubit LME states with entangling isometries $V_{k}: \mathbb C^2 \to \mathbb C^2 \otimes \mathbb C^2$ [cf.~\cref{eq:compressed_CJ}] have the form \cref{eq:LME_phase} up to LU. 
\end{acorollary}

\end{document}